\documentclass{jpp}
\usepackage{graphicx}
\usepackage{epstopdf, epsfig}

\usepackage[dvipsnames]{color}

\newcommand{\kmpskpc}{km~s\ensuremath{^{-1}}~kpc\ensuremath{^{-1}}}
\newcommand{\kmps}{km~s\ensuremath{^{-1} }}

\shorttitle{Simulated observations of the outer galaxy disks}
\shortauthor{S.A.Khoperskov and G. Bertin}

\title{Synthetic HI observations of spiral structure in the outer disk in galaxies}

\author{Sergey A. Khoperskov\aff{1,2,3}
 \corresp{\email{sergey.khoperskov@unimi.it}},
 Giuseppe Bertin\aff{1}}

\affiliation{\aff{1}Dipartimento di Fisica, Universit\`{a} degli Studi di Milano, via Celoria 16, I-20133 Milano, Italy
\aff{2}Institute of Astronomy, Russian Academy of Sciences, Pyatnitskaya st., 48, 119017 Moscow, Russia
\aff{3}Sternberg Astronomical Institute, Moscow M.V. Lomonosov State University, Universitetskij pr., 13, 119992 Moscow, Russia}

\begin{document}

\maketitle

\begin{abstract}
By means of 3D hydrodynamical simulations, in a separate paper we have discussed the properties of non-axisymmetric density wave trains in the outermost regions of galaxy disks, based on the picture that self-excited global spiral modes in the bright optical stellar disk are accompanied by low-amplitude short trailing wave signals outside corotation; in the gas, such wave trains can penetrate through the outer Lindblad resonance and propagate outwards, forming prominent spiral patterns. In this paper we present the synthetic 21~cm velocity maps expected from simulated models of the outer gaseous disk, focusing on the case when the disk is dominated by a two-armed spiral pattern, but considering also other more complex situations. We discuss some aspects of the spiral pattern in the gaseous periphery of galaxy disks noted in our simulations that might be interesting to compare with specific observed cases.
\end{abstract}

\section{Introduction}

Galaxies are macroscopic systems in which many complex collective phenomena are governed by the long-range gravity force. In this respect, one often refers to self-gravitating systems as to gravitational plasmas. Spiral structure in galaxy disks is a classic collective phenomenon that has attracted the interest of astrophysicists, dynamicists, and plasma physicists over the last century, starting with the pioneering work of Jeans, Chandrasekhar, and Oort.

In a number of galaxies (e.g., in NGC~6946 and NGC~1512/1510), deep HI and UV observations  indicate the presence of well-developed gaseous spiral arms in the regions outside the bright optical disk~\citep{2008A&A...490..555B,2009MNRAS.400.1749K}. Such prominent structures in regions where the disk is light and the stars are not the dominant visible component may be puzzling. Some observations detect ongoing star formation in the outer regions of galaxies~\citep[e.g., see][]{2010AJ....140.1194B,2015MNRAS.451.4622Y}. It is thus important to study what mechanisms may operate to support the observed structures, which might also have some consequences on the general problem of dark matter in disk galaxies. In general, investigations of the galactic outskirts is an important tool for understanding the chemical evolution of galaxies, the problem of radial mixing, and the growth of galaxies in the cosmological context~\citep{2012A&A...548A.126M}.

It is commonly thought that grand-design spiral structure in galaxies is associated with density waves. In one scenario worked out in quantitative detail, it is argued that grand-design spiral structure results from the superposition of few, self-excited, long-lived global modes~\citep[][see also ~\citealt{2014dyga.book.....B} where alternative scenarios are also reviewed and many up-to date references are provided]{1989ApJ...338...78B,1989ApJ...338..104B,1996ssgd.book.....B}. Many $N$-body and hydrodynamical simulations have addressed the viability of such scenario~\citep{1998ApJ...504..945L,2000ApJ...541..565K,2007A&A...473...31K,2012MNRAS.427.1983K}. However, little discussion has been provided of the physical nature of  structures in the outskirts of galaxies. Under the hypothesis that the light outer disk would be by itself unable to generate prominent arms, several mechanisms have thus been considered as a source of spiral patterns; in particular, attention has been focused on tidal events~\citep{1987MNRAS.228..635N,2005ApJ...635..931B,2009ApJ...703.2068D} and on deviations from axisymmetry in the dark matter distribution~\citep{2012ARep...56...16K,2014AJ....147...27V}. 

In this paper we follow the picture that the structures observed in the outer disk are the manifestation of the density waves that, by outward transport of angular momentum, are expected to be associated with the excitation of the spiral modes that dominate the bright optical disk. That is, we consider a hydrodynamical model of the galactic outer disk subject to perturbations beyond the outer Lindblad resonance that are described approximately by the linear analysis of~\citet{2010A&A...512A..17B}.

The structure of this article is as follows. In Sects.~2 and 3 we briefly summarize the general set-up and results of a separate paper~\citep{2015MNRAS.451.2889K} where the present dynamical scenario is described in full detail. In particular, the dynamical model and numerical approach are summarized in Sect.~2, whereas in Sect.~3 we briefly illustrate the evolution of the outer gaseous disk when a single-mode perturbation is imposed at the inner boundary and in other more complex cases.  The remaining sections contain the key results of the present paper. In Sect. 4 we briefly describe the approach used for the calculation of the 21-cm spectra. In Sect.~5 we illustrate the synthetic spectra that would be observed in HI (21~cm line) if the disk behaves as predicted by our simulations. We also consider the role of the spiral pattern on the shape of the synthetic velocity maps and line-of-sight~(LOS) velocity distribution. A discussion of the main aspects of these results and other points of observational interest is given in Sect.~6.

\section{Model}
For the numerical simulations of the gaseous disk evolution, we make use of the TVD MUSCL~(Total Variation Diminishing Multi Upstream Scheme for Conservation Laws) scheme~\citep[for a description of the structure of the code, see ][]{2014JPhCS.510a2011K}. Below we show the results of 3D~simulations on a uniform $2048\times2048\times128$ Cartesian grid. The computational box size is $144\times144\times9$~kpc, which corresponds to a spatial resolution of $\approx 70$~pc.

\begin{figure}
\begin{minipage}[h]{0.35\linewidth}
\includegraphics[width=1\linewidth]{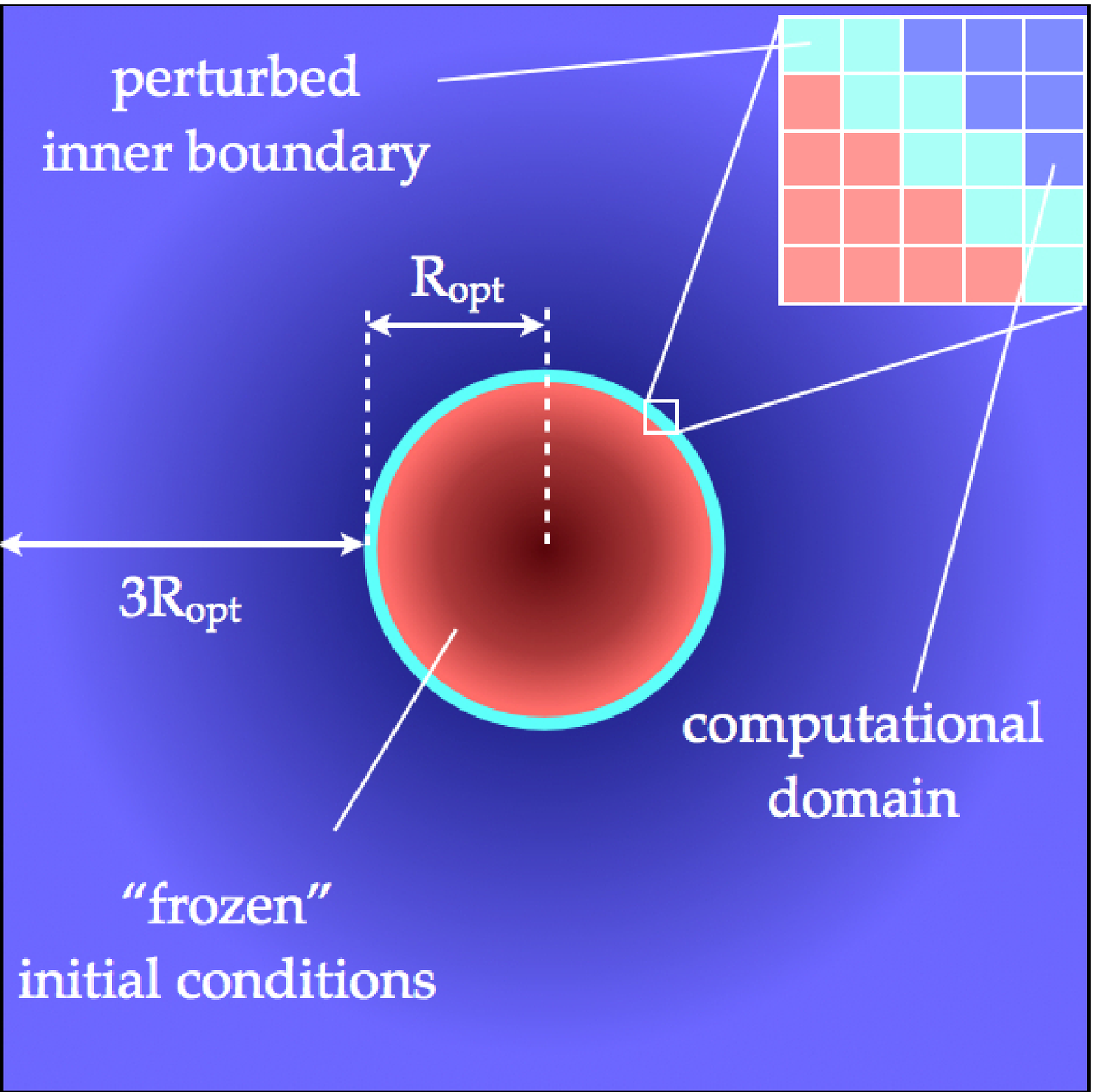}
\end{minipage}
\hfill
\begin{minipage}[h]{0.65\linewidth}
\includegraphics[width=0.32\hsize]{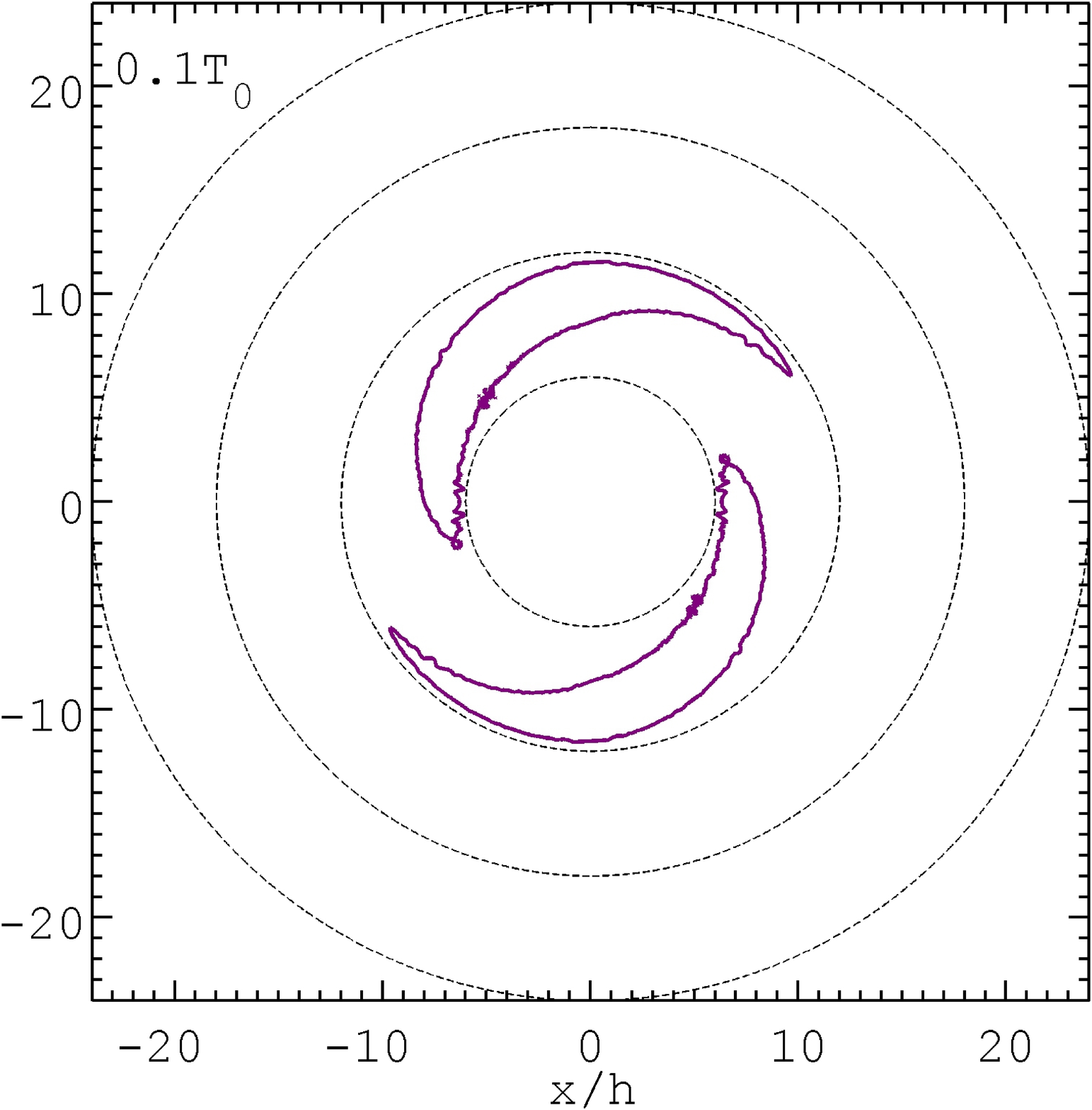}\includegraphics[width=0.32\hsize]{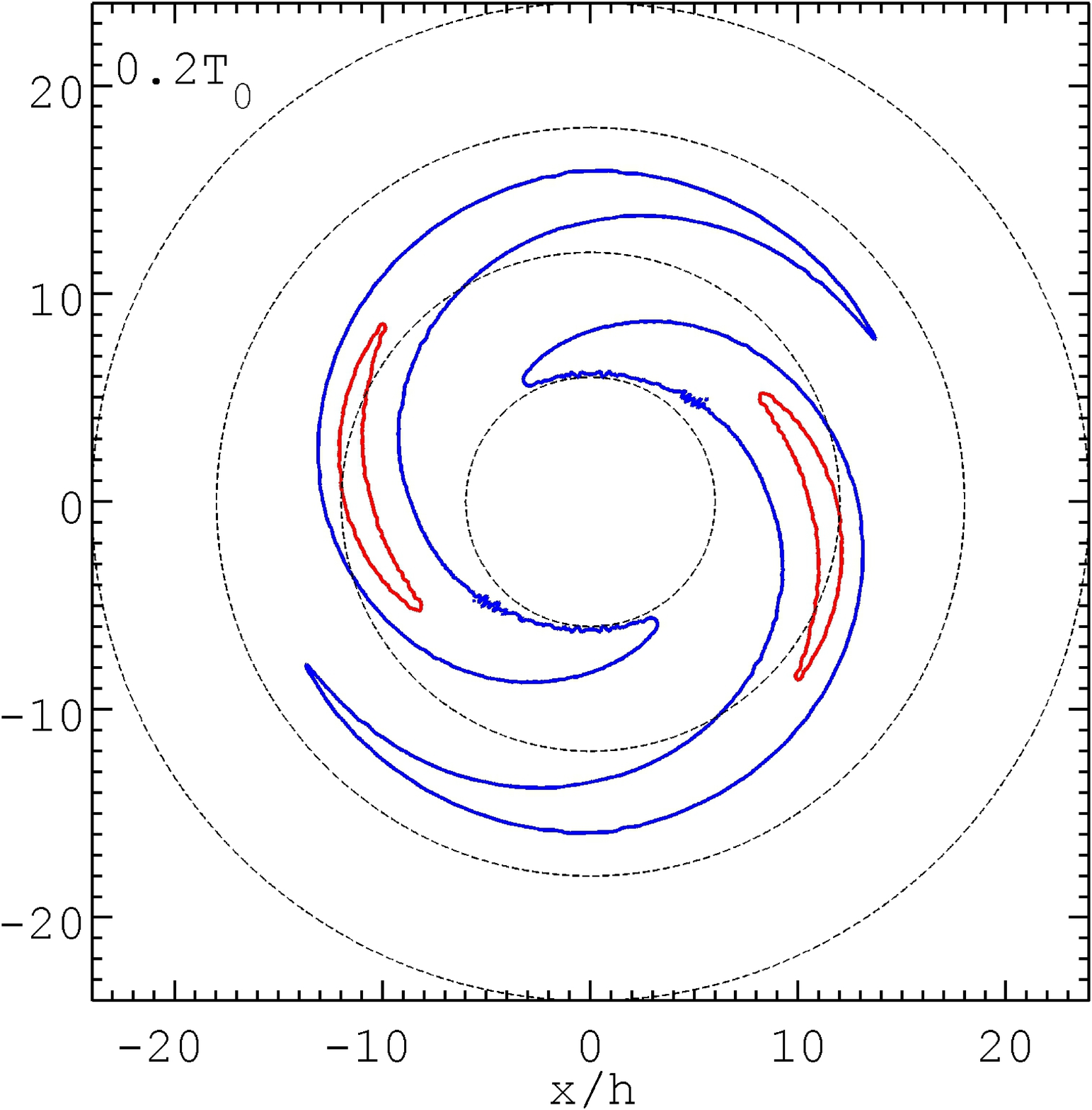}\includegraphics[width=0.32\hsize]{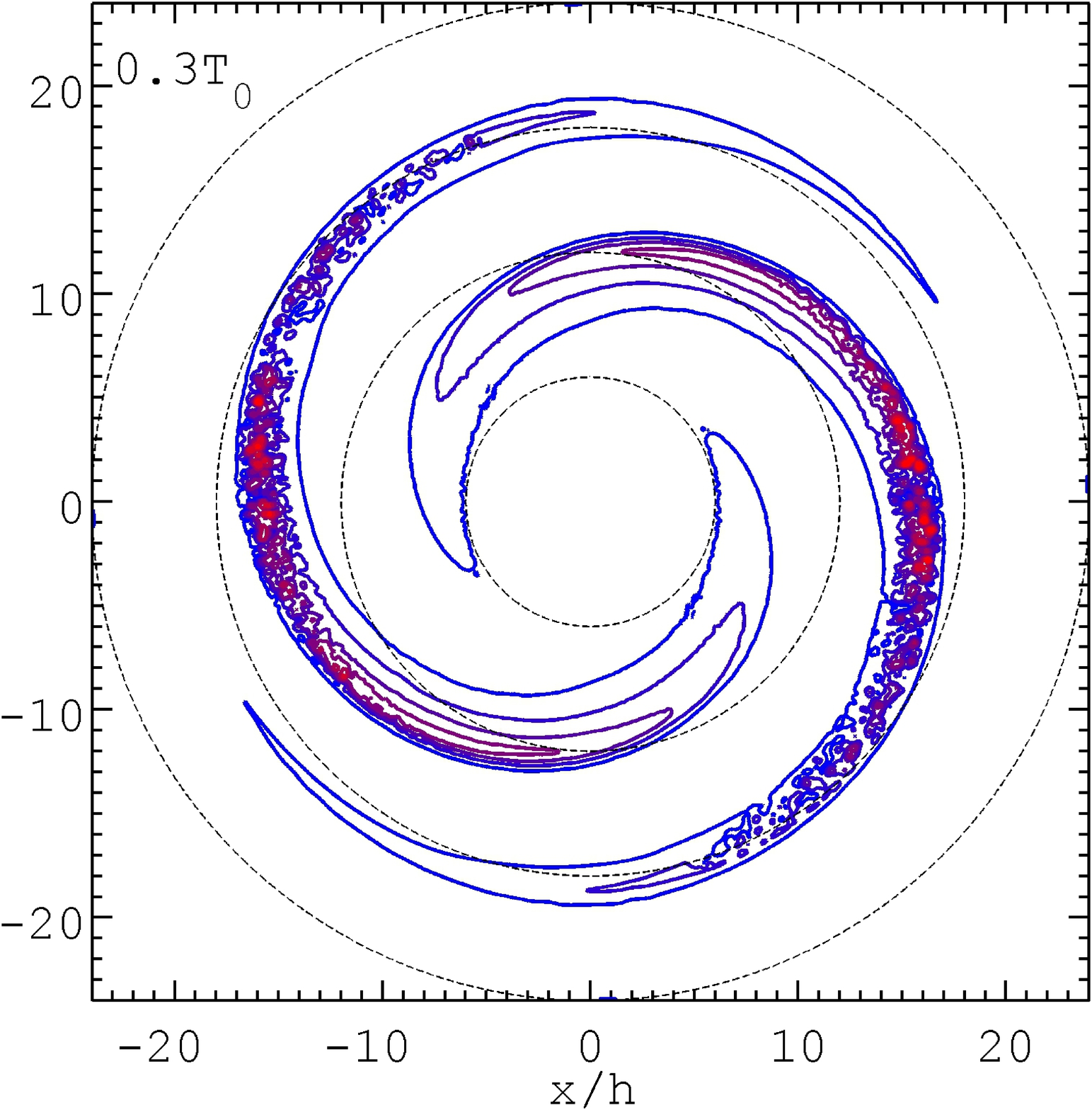} \\\includegraphics[width=0.32\hsize]{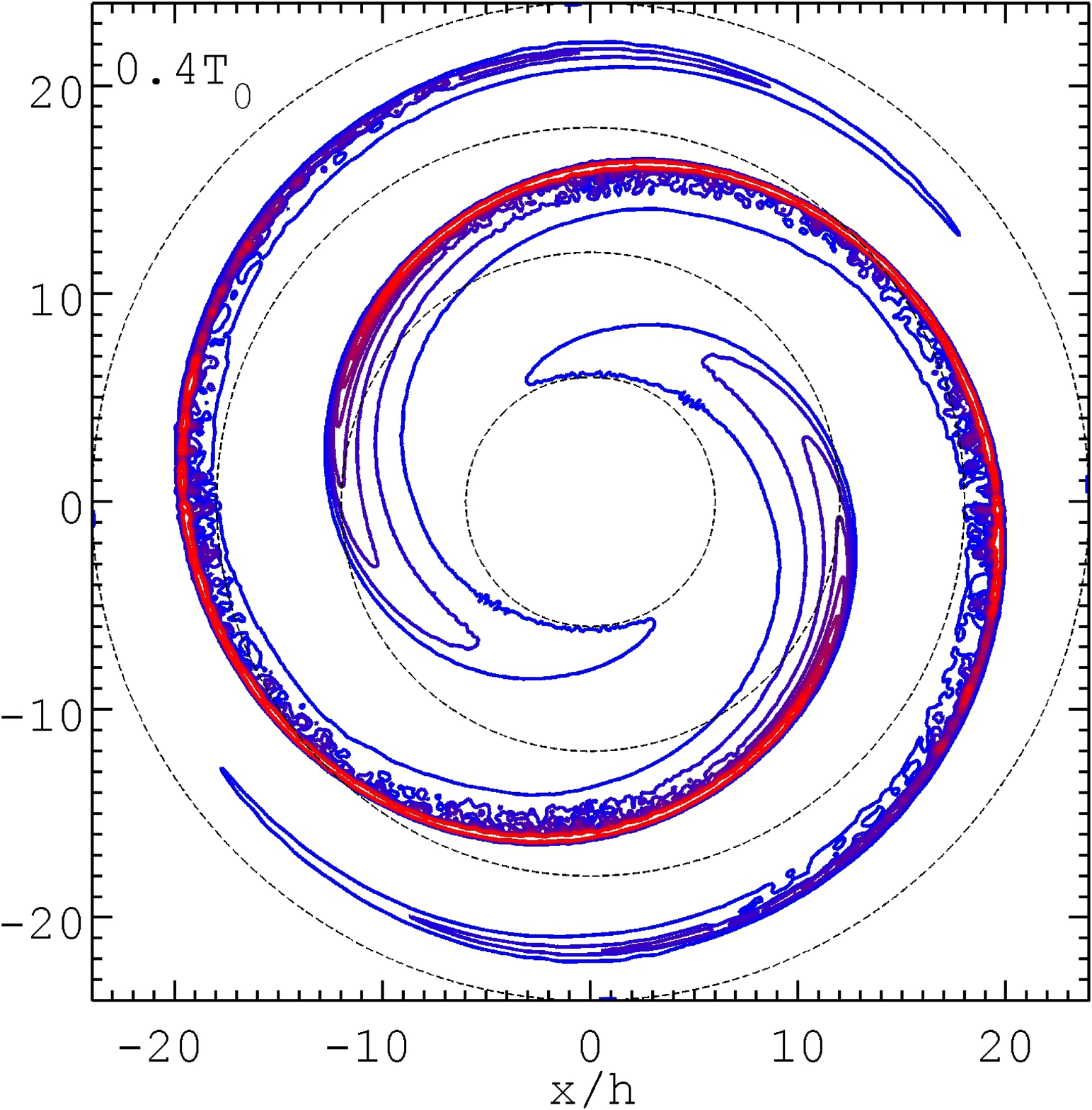}\includegraphics[width=0.32\hsize]{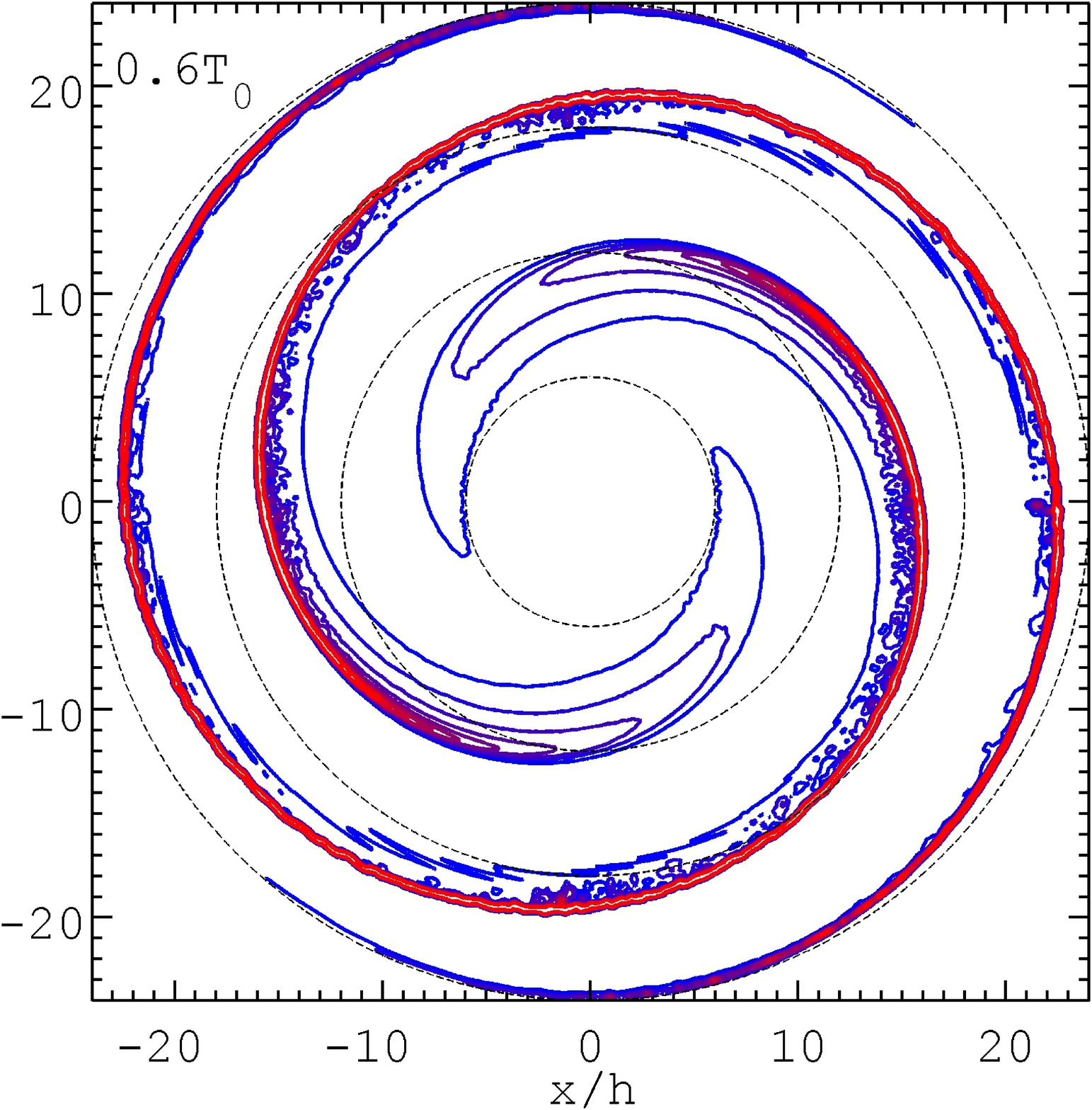}\includegraphics[width=0.32\hsize]{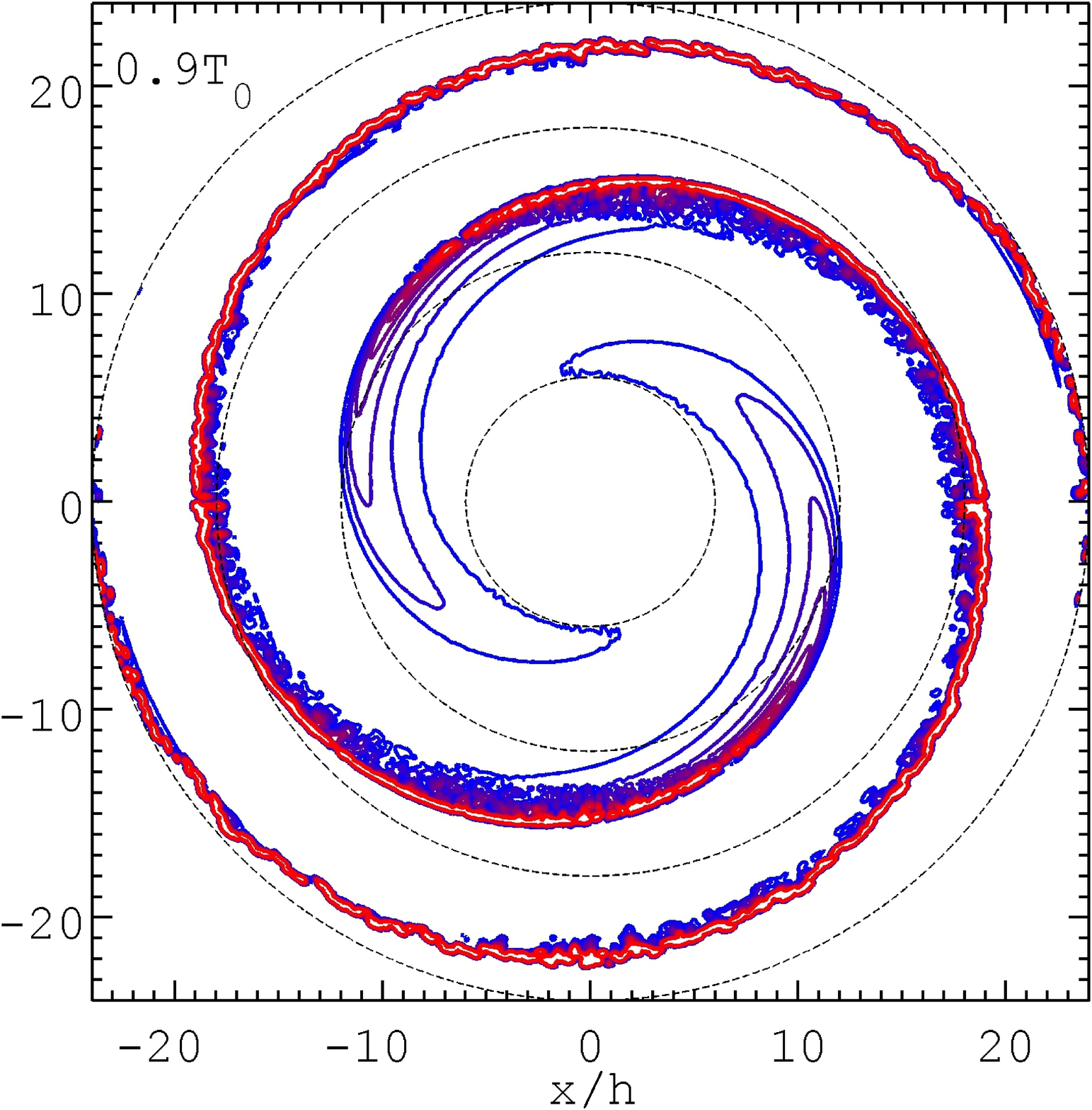}
\end{minipage}
\caption{The left panel \citep[taken from][]{2015MNRAS.451.2889K} shows the general scheme of the computational model. Here the inner red disk,  inside $r=6h$, is an area of ``frozen" initial conditions. The thin, light circular annulus (cyan line) is the site where the outgoing density-wave perturbation is imposed.  The right panel shows the evolution of the relative surface density perturbation (positive part of the quantity $\Sigma_1/\langle\Sigma_g \rangle$). Contours vary from blue to red, which corresponds to values of the relative density perturbation in the range $0.01-1$ with steps of $0.05$.  The plots are drawn in the inertial, nonrotating frame of reference. The time scale is the rotation period $T_0 \approx 0.5$~Gyr at $6h = 18$~kpc. Black circles are drawn at radii $6h$, $12h$, $18h$, and $24h$.}\label{fig::fig1}
\end{figure}

We calculate the evolution of a self-gravitating gaseous disk embedded in a fixed external potential generated by a quasi-isothermal dark-matter halo and by an exponential stellar disk. The parameters of the external potential are chosen so as to support a flat rotation curve in the outer parts, at $\approx 210$~\kmps. The vertical structure of the gas disk is established by the conditions of hydrostatic equilibrium. We then start the computation by imposing a spiral density-wave perturbation in an annulus at $r=6h$, which we call ``the inner boundary"~(see Fig.~\ref{fig::fig1}, left panel; for simplicity, we indicate the annulus by $R_{\rm opt}$); here $h$ denotes the radial scale length of the exponential stellar disk. The density wave represents the outgoing signal coming from the bright optical disk and carrying angular momentum outwards; spiral structure in the main disk is taken to be self-excited, as a result of such transport. The dynamics of the inner disk, inside the annulus, is not treated in the simulation and is only assumed to provide the adopted inner boundary conditions. 

Basically, we consider perturbations of the relevant hydrodynamical quantities ${\bf \hat{X}}$ at the inner boundary of the form:
\begin{equation}
{\bf \hat{X}} \propto \cos\left(m\theta- \omega t\right)\,, \label{eq::perturb}
\end{equation}
where  $m$ is the mode azimuthal number, $\Omega_p=\omega/m$ is the angular speed of the spiral density perturbation, $t$ is the time, $\theta$ is the angular coordinate. The relative amplitudes of the perturbation for pressure and velocity (for the short-trailing wave-branch) can be found straightforwardly from Eqs. (7), (8), (9), and (13) in the paper by~\cite{2010A&A...512A..17B}. Here we focus on one model~(B1) in which the imposed perturbation is characterized by a single $m=2$ mode. Examples of the behavior of the outer disk in the presence of more complex structures are given below in this paper and in~\citet{2015MNRAS.451.2889K}. This is a simple representative case to demonstrate the general properties of the evolution of the outer gaseous disk.  At the inner boundary the relative amplitude of the density wave is $\Sigma_1/\langle\Sigma_g \rangle = 0.1$. The pattern speed of the mode is chosen to be $30$~\kmpskpc, which implies that the outer Lindblad resonance~(OLR) is located at $5.6$~h ($h=3$~kpc is the adopted stellar disk scale length).  Thus OLR is inside the inner boundary. In the disk inside the circle at $r=6h$, the gas distribution is kept ``frozen" and axisymmetric~(see left panel in Fig.~\ref{fig::fig1}). The blue area outside, beyond the circular annulus, is the actual computational domain, from $6h$ out to $24h$, where the hydrodynamical quantities evolve.

\section{Establishment of spiral structure in the outer disk}
For the simple case studied here, the evolution of the gas surface density perturbation is illustrated in the right panel of Fig.~\ref{fig::fig1}. The two-armed trailing spiral structure imposed at the inner boundary propagates outwards and its amplitude increases rapidly with radius. At time $\approx T_0$ after the beginning of the simulation, a quasi-stationary structure sets in. We found that the large-scale structure of the outer disk does not change significantly for $t>T_0$, for several dynamical times. Surely, the small-scale structures do evolve on a relatively small time scale, but this does not affect the large-scale morphology of the spiral pattern. That is why we do not show further frames in the figure. The spiral pattern thus established in the outer regions rotates rigidly with the imposed angular velocity $\Omega_p=30$~\kmpskpc.

\begin{figure}
\includegraphics[width=0.495\hsize]{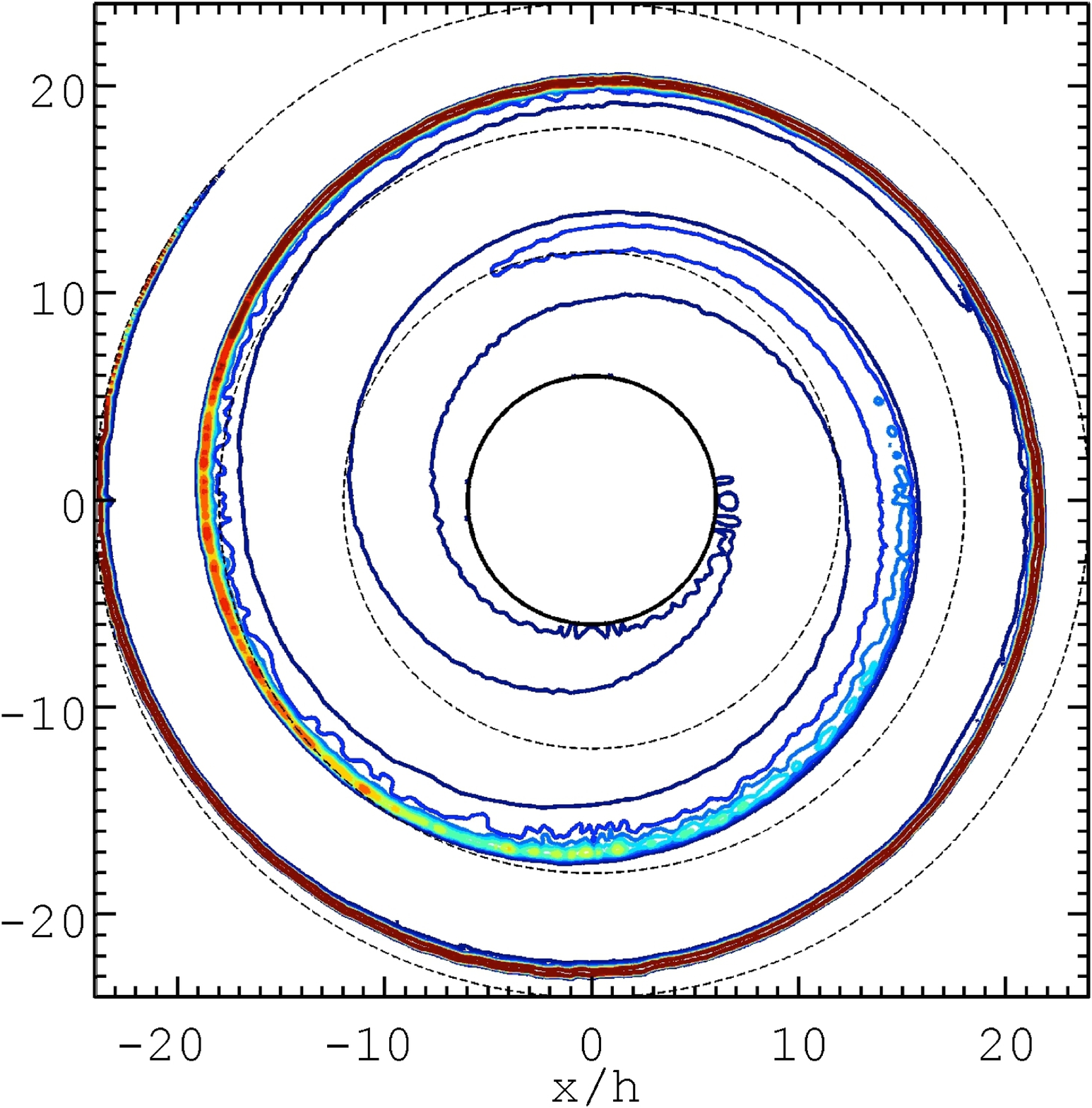}\includegraphics[width=0.495\hsize]{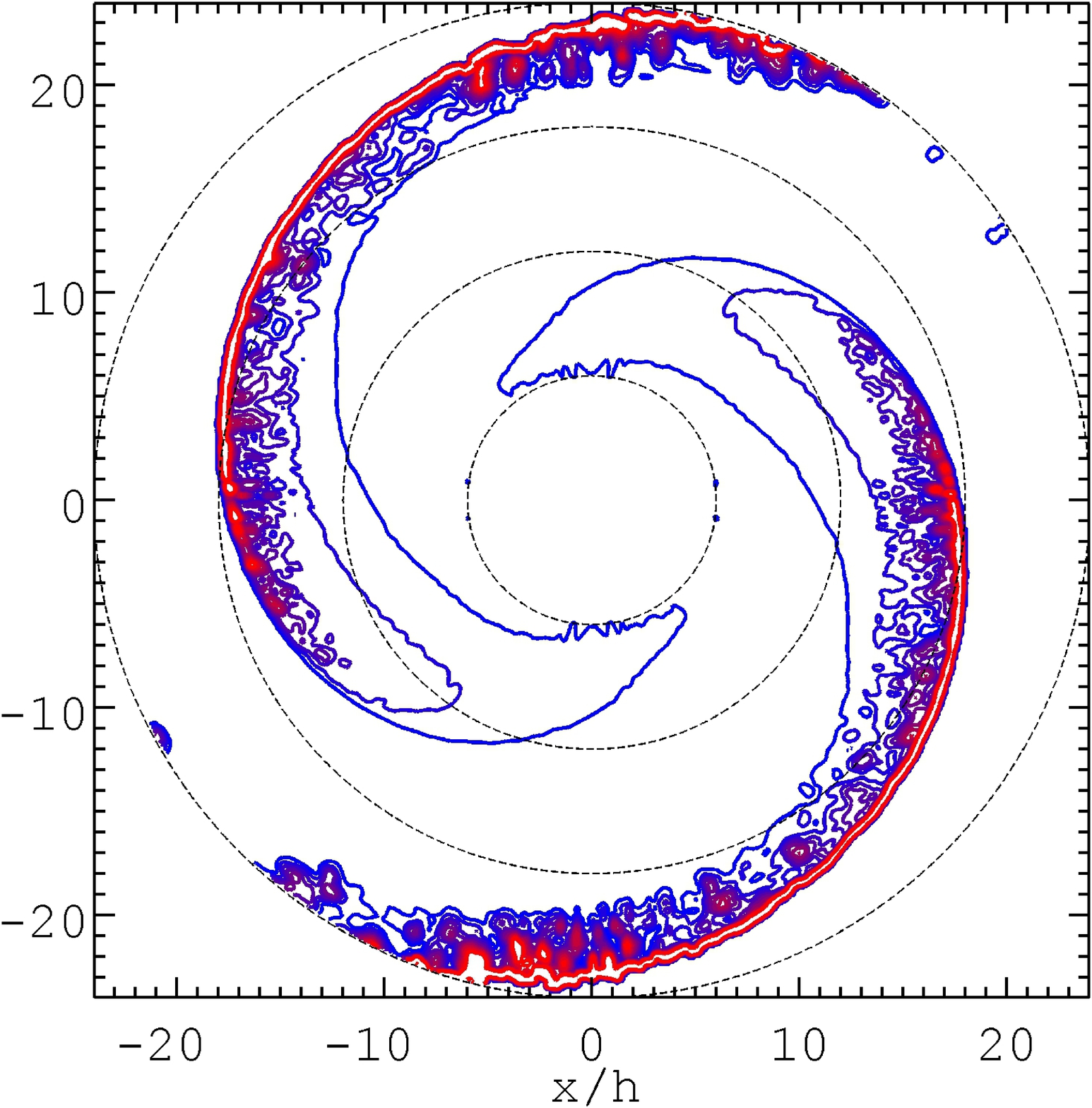}\newline\includegraphics[width=0.495\hsize]{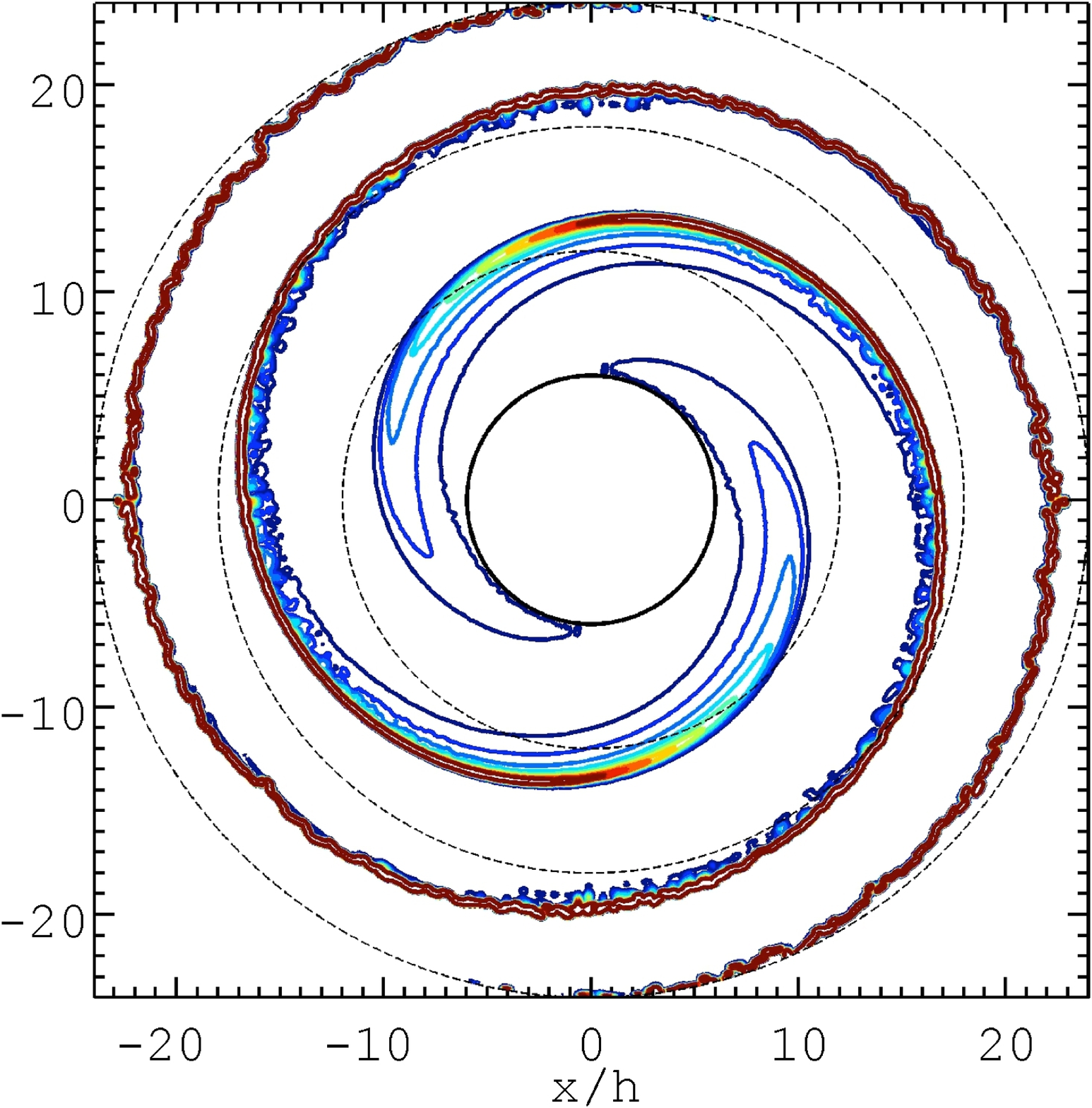} \includegraphics[width=0.495\hsize]{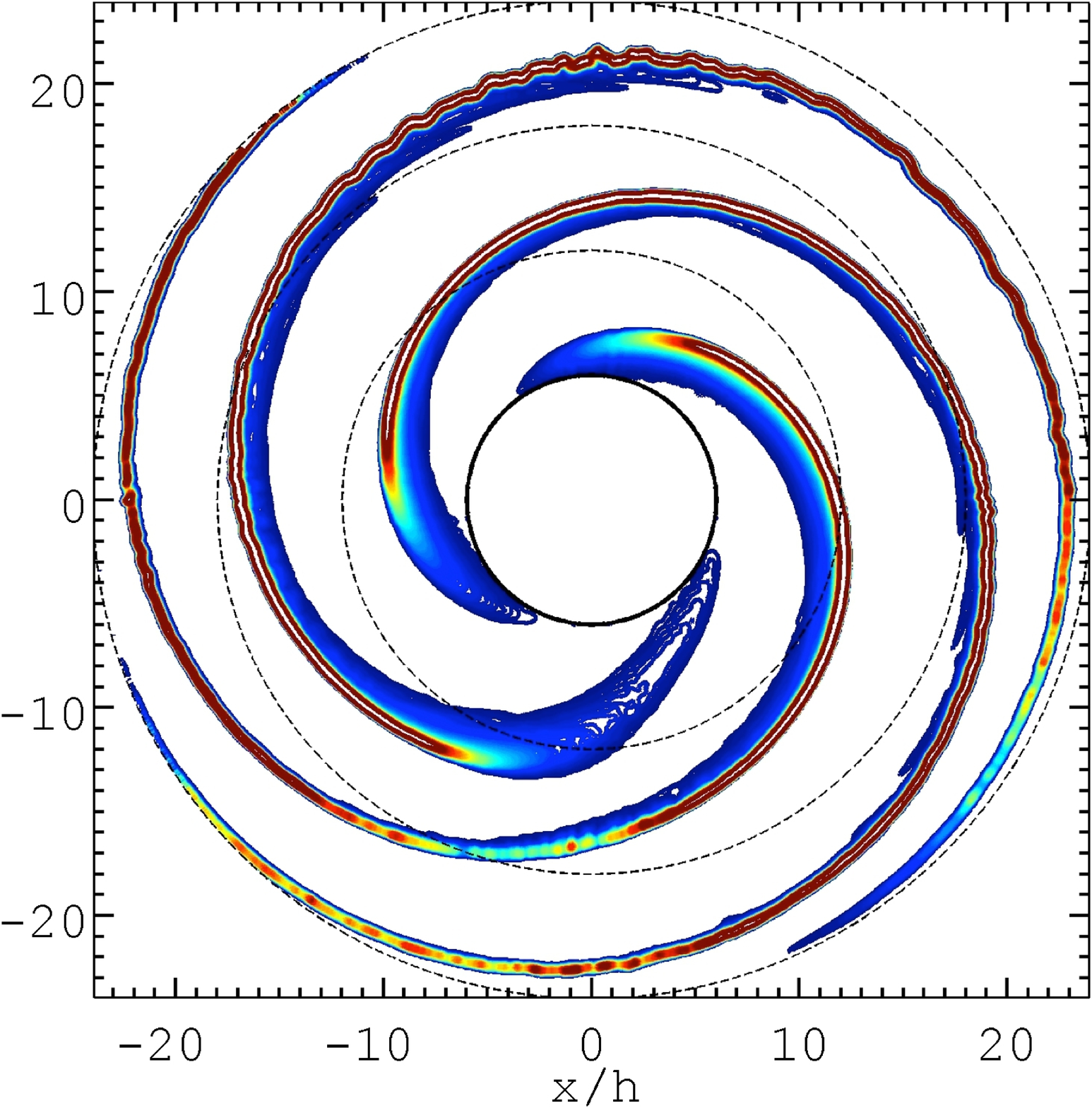}
\caption{The relative surface density perturbations (positive part of the quantity $\Sigma_1/\langle\Sigma_g \rangle$) for models at time $t\approx T_0$, with different imposed perturbations. Top row: single $m=1$ mode~(left,  model E1) and single $m=2$ mode on a gaseous disk characterized by higher velocity dispersion~(right,  model B7). Bottom row: morphology produced by a single $m=2$ mode~(left,  model B1) and by the superposition of three modes with $m=1$, $m=2$, and $m=3$~(right,  model J1). Contours vary from blue to red, which corresponds to values of the relative density perturbation in the range $0.01-1$.  The plots are drawn in the inertial, nonrotating frame of reference. Black circles are drawn at radii $6h$, $12h$, $18h$, and $24h$.}\label{fig::fig2}
\end{figure}
Close to the inner boundary, inside the disk of radius $\approx 12h$, the spiral structure is regular and characterized by a low relative amplitude $\leqslant 0.1-0.3$. Here the spiral pattern follows the linear theory described by~\citet{2010A&A...512A..17B}. Further out, the wave amplitude increases rapidly; the shape of the density perturbation departs from being sinusoidal and the density profile of the perturbation along a given circle becomes asymmetric. At radii beyond $\approx 12-15 h$, rather narrow shocks form because of the supersonic motion of the pattern across the disk. The shocks are unstable with respect to the wiggle instability, which develops behind the shock front because of the strong shear flow~\citep[see description of the instability in][]{2004MNRAS.349..270W}. 

The numerical simulations that we have performed suggest that any type of short trailing density perturbation, which in the gas can penetrate through the outer Lindblad resonance, produces an extended spiral pattern.  Examples of extended patterns outside the bright optical disk are shown in Fig.~\ref{fig::fig2}~(some key parameters of the models are listed in Table~\ref{tab::tab1}). Instabilities associated with the shear flow are likely to be the major driver of spur formation in the vicinity of the spiral shocks. This point is also supported by simulations run on gaseous disks characterized by higher velocity dispersion where more open spirals are formed~(see model B7 in Fig.~\ref{fig::fig2}). These configurations are more unstable and smaller-scale structures arise. Independent fluid simulations by~\citet{2004MNRAS.349..270W} also suggest that open spiral-shocks should be more unstable. However, the detailed mechanism of the origin of the wiggle instability is currently under debate; in particular it is not clear whether a relatively standard Kelvin-Helmholtz instability or some other qualitatively different mechanism is actually at work~\citep[in particular, see][]{2014ApJ...789...68K,2015arXiv150607178K}. 

\begin{table*}
\begin{center}
\caption{ Parameters of different runs. Here $m$ is the number of arms and $A_0$ is the relative amplitude of the initially imposed density perturbation. For model J1, the  inner boundary is perturbed with the superposition of three modes. The last column lists the value of the adopted gas velocity dispersion in the unperturbed outer disk.}\label{tab::tab1}
\begin{tabular}{ccccccc}
\hline
Run   & $m$ & $A_0$ & $\sigma_{\rm th}$  \\
   &         &          &  \kmps \\
\hline
E1      & 1  	  & 0.05 			 & 3.4 \\ 
B1      & 2  	  & 0.1  			 & 3.4 \\ 
B7      & 2  	  & 0.1 			     & 5 	\\ 
J1      & 1, 2, 3  & 0.05, 0.1, 0.15  & 3.4  \\ 
\hline
\end{tabular}
\end{center}
\end{table*}

\section{HI spectra calculation}

To compare the properties of the gaseous disk as determined by the numerical simulations with the observations, we calculate the synthetic spectra for the atomic hydrogen emission in the 21~cm line. The line profile is mostly determined by turbulence (and other local motions) in the gas and by instrumental beam smearing, whereas the peak velocity is mostly determined by the gas macroscopic motions (rotation and systematic deviations associated with density waves). Thus from the simulated spectra we can identify the general rotation curve of the galaxy model and the impact of spiral structure on the large-scale kinematics of the gas. 

To do so, we use a standard technique for the calculation of the brightness of an inclined galactic disk~\citep{1954BAN....12..117V, wilson2009tools}, assuming an inclination of $30^\circ$. Namely, we calculate the distribution of the brightness temperature in the velocity range $[-250; 250]$~\kmps in each pixel of a projected image. 
 We can express the optical thickness $\tau_v$ of a gaseous element with line-of-sight velocity in the range ($v,v+\delta v$) as:
\begin{equation}
\tau_v = \frac{3hcA_{10}}{32\pi\sqrt{2\pi}} \int\limits_s \frac{N_H}{\sigma_v k_{\rm B}T} \exp\left[-\frac{(v-u)^2}{2 \sigma_v^2 }\right] du\,,\label{eq::radtr9}
\end{equation}
where from the simulated data we know the distribution of the LOS velocity $u(s)$, $N_H$ is the gas concentration,  $\sigma_v$ is the gas velocity dispersion, $s$ is the coordinate along the line of sight, $T$ is the gas temperature, $k_{\rm B}$ is the Boltzmann constant, $h$ is the Planck constant, $c$ is the speed of light, $A_{\rm 10}$ is the Einstein coefficient for spontaneous emission in a two-level system.

 In the calculations we assume that $\sigma_v$  is the superposition of the ``thermal contribution" of the velocity dispersion and of the random motions of the gas along the line of sight:
\begin{equation}
\sigma^2_v = \sigma^2_{\rm th} + \sigma^2_{\rm random}.
\end{equation}

 We recall that in our simulations the ``thermal contribution" $\sigma_{\rm th}$ is dominated by an effective thermal speed associated with the turbulent velocity dispersion present in the unperturbed gas; this velocity dispersion was called $c$ in Eq.(2) of the paper by~\citet{2015MNRAS.451.2889K}. Integrating the second moment of the velocity distribution along the line of sight gives what we call random component of the gas velocity dispersion $\sigma_{\rm random}$.

In dealing with the radiative transfer equations, we assume that the gas density is sufficiently high and that for the 21 cm line transition $h\nu_0/k_{\rm B}T\ll 1$.

In the following section we apply this procedure in the context of the extended spiral structure found in our hydrodynamical simulations. We consider two models: one model with a single $m=2$ pattern and developed small-scale spurs  (model B1) and one model with the superposition of several modes ($m=1,2,3$) imposed at the inner boundary  (model J1).

\section{Synthetic observations of the simulated galaxies}

\begin{figure}
\includegraphics[width=0.995\hsize]{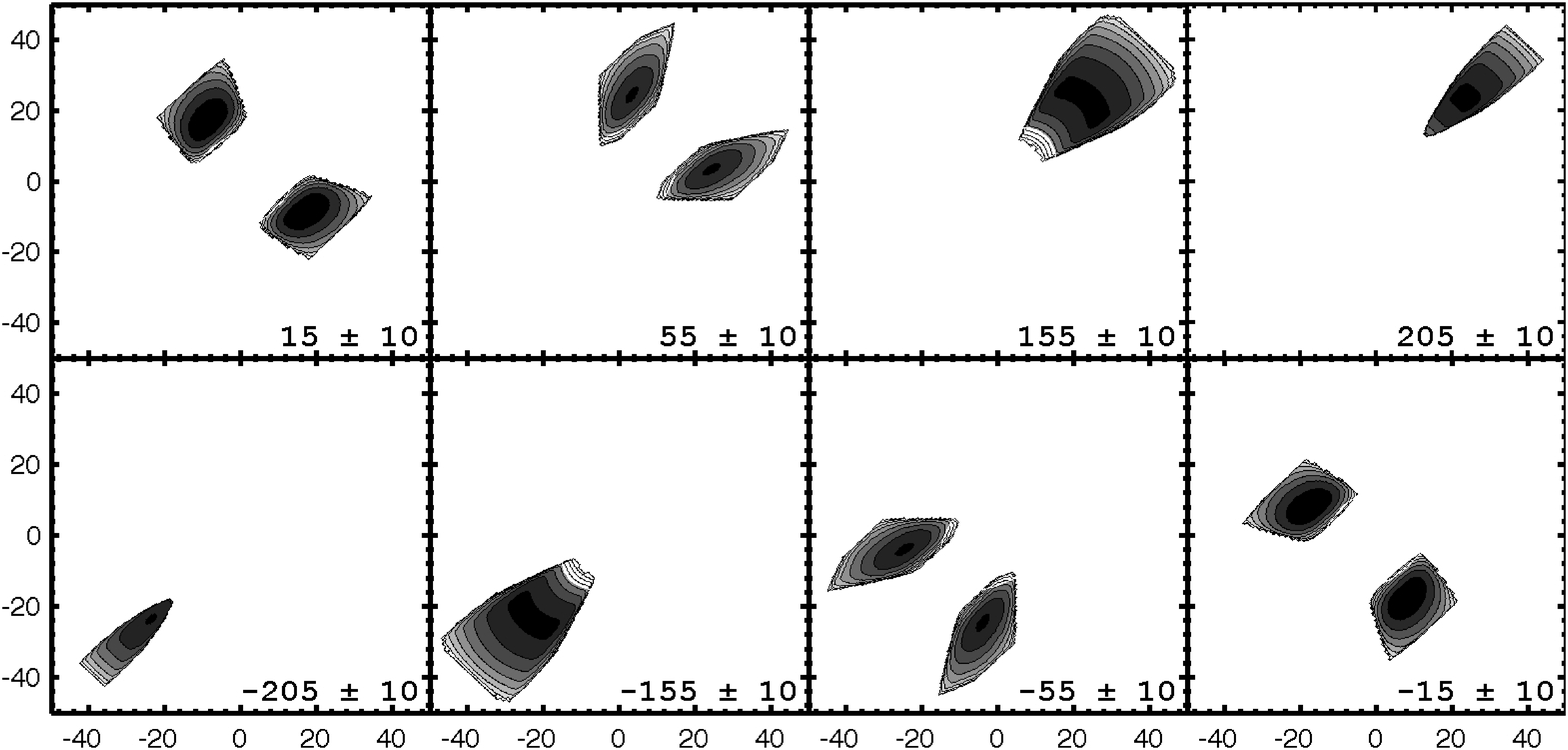}\hfill \includegraphics[width=0.995\hsize]{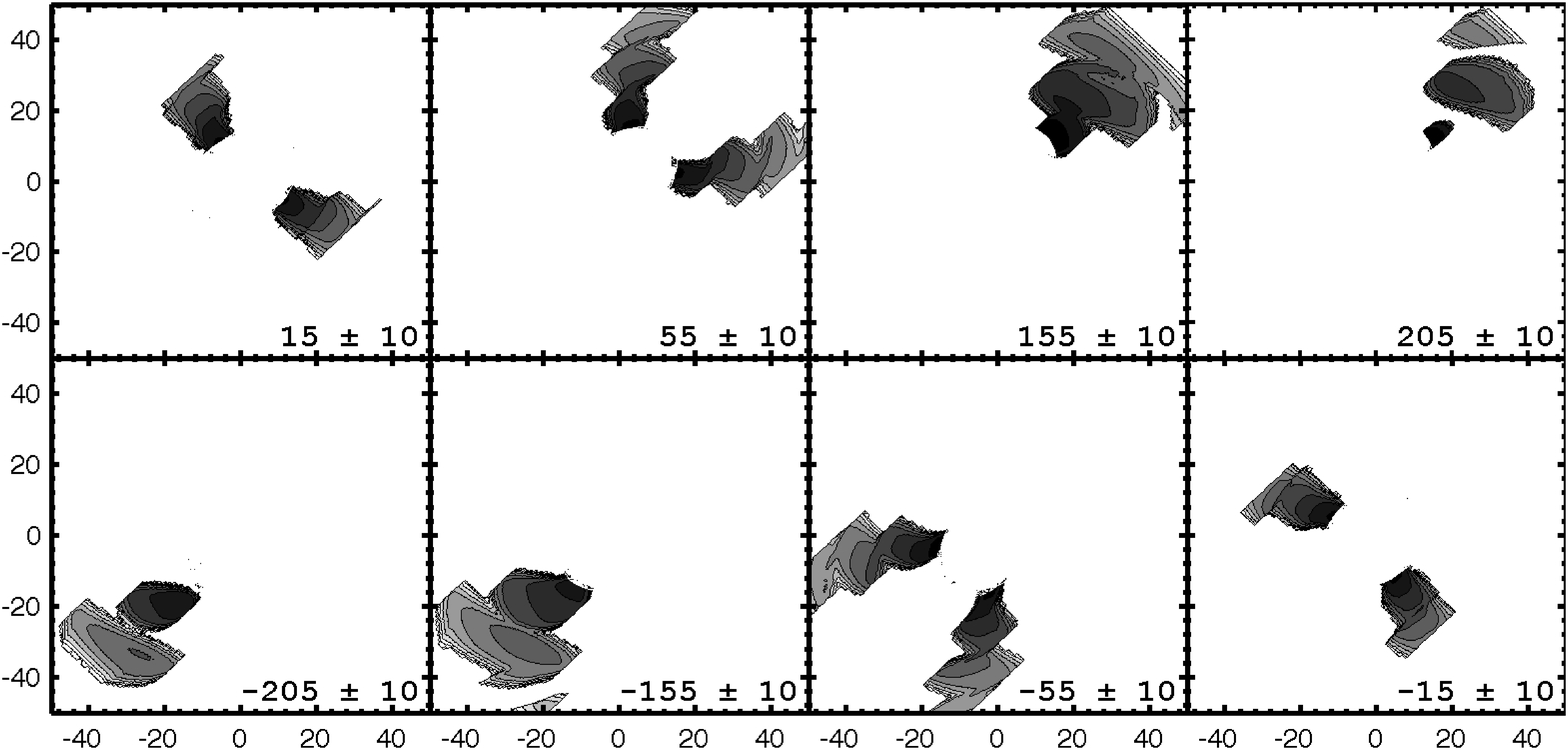}
\caption{Synthetic HI flux distribution at varying line-of-sight velocities for a simulated gaseous disk inclined at an angle of $30^\circ$ for the  model J1 with the superposition of several modes ($m=1,2,3$). Each frame corresponds to a different velocity channel; the value of the velocity is shown on the bottom right corner of the frames.  Contours represent HI column density levels of 0.5, 1, 2, 3, 4, 5, 10, and 15 in units of $10^{19}$~cm$^{-2}$. The top panel shows the unperturbed initial state; the bottom  panel illustrates the properties of the gas flow at $t=0.9T_0$. The area within the inner boundary size $r = 6h = 18$~kpc is masked, because in that region the disk is kept fixed in the simulations.}\label{fig::hidatacub}
\end{figure}

\begin{figure}
\includegraphics[width=0.995\hsize]{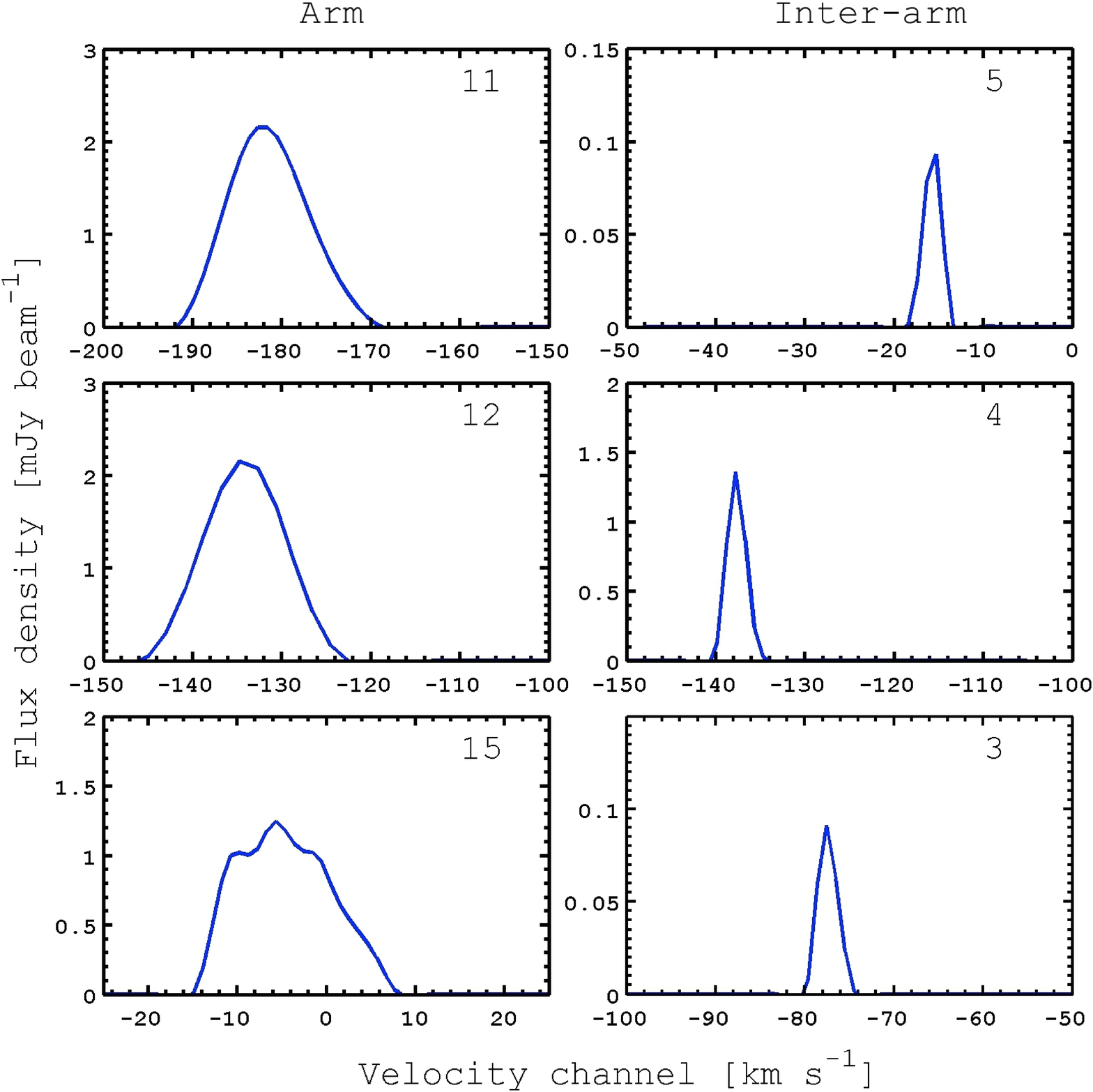}
\caption{  Line profiles extracted from the HI data
cube for a simulated gaseous disk  inclined at an angle of $30^\circ$ for the model J1 with the superposition of several modes ($m=1,2,3$). Left frames show line profiles in the vicinity of spiral arms, right frames refer to the inter-arm regions. The estimated FWHM value is indicated in the top-right corner of each panel in units of \kmps.}\label{fig::line_profiles}
\end{figure}

\begin{figure}
\includegraphics[width=0.47\hsize]{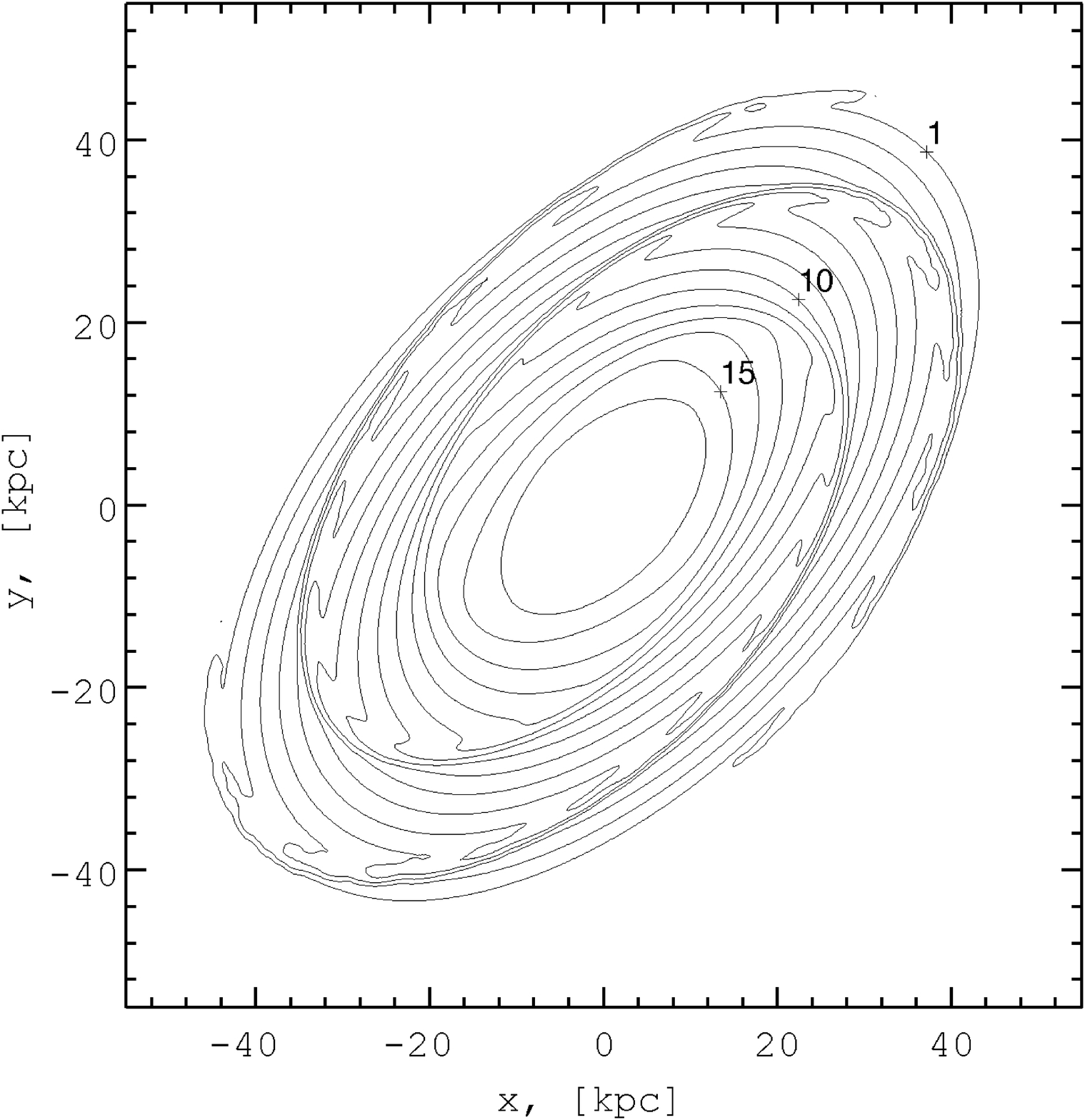}
\includegraphics[width=0.47\hsize]{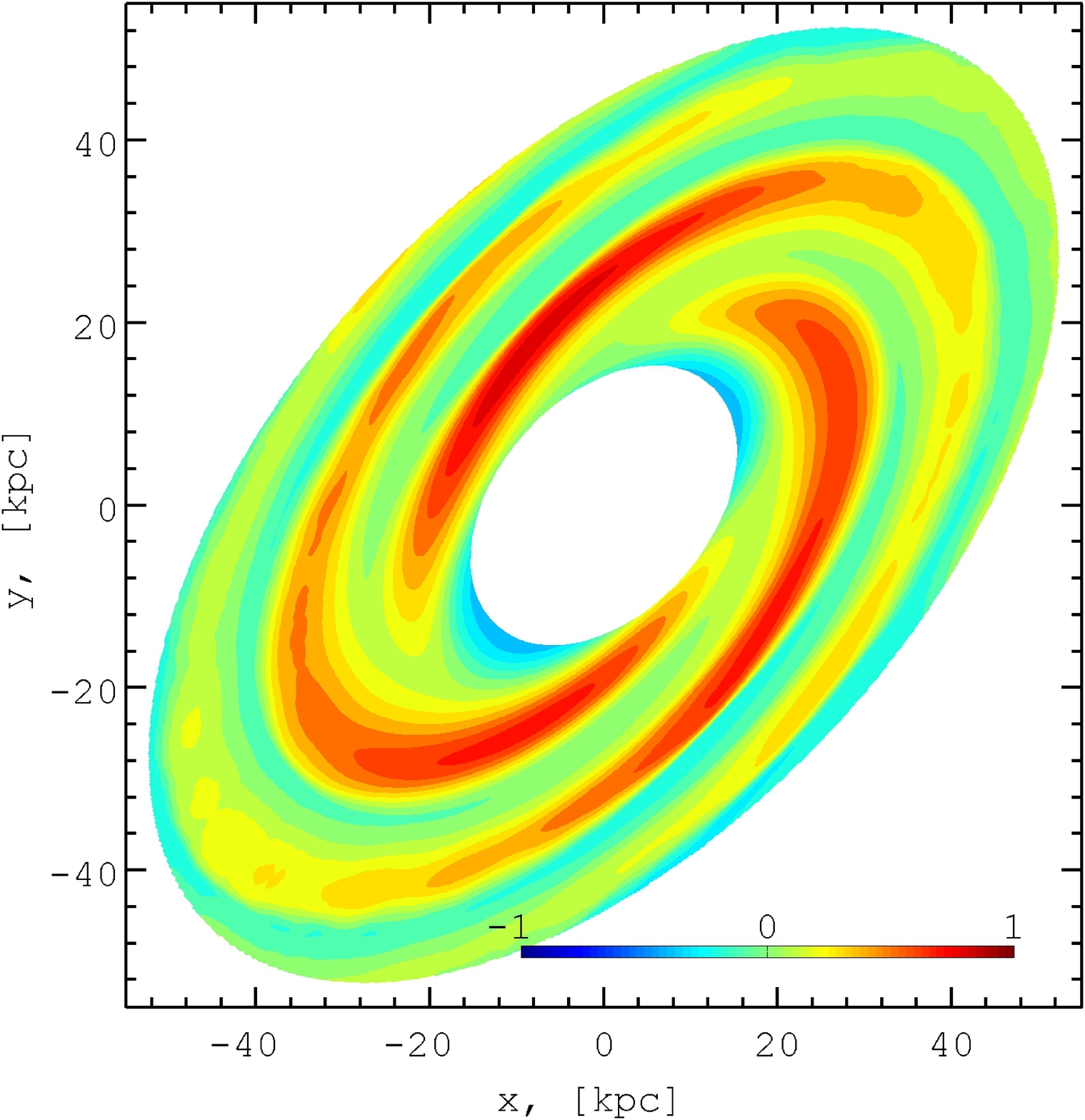} 
\includegraphics[width=0.47\hsize]{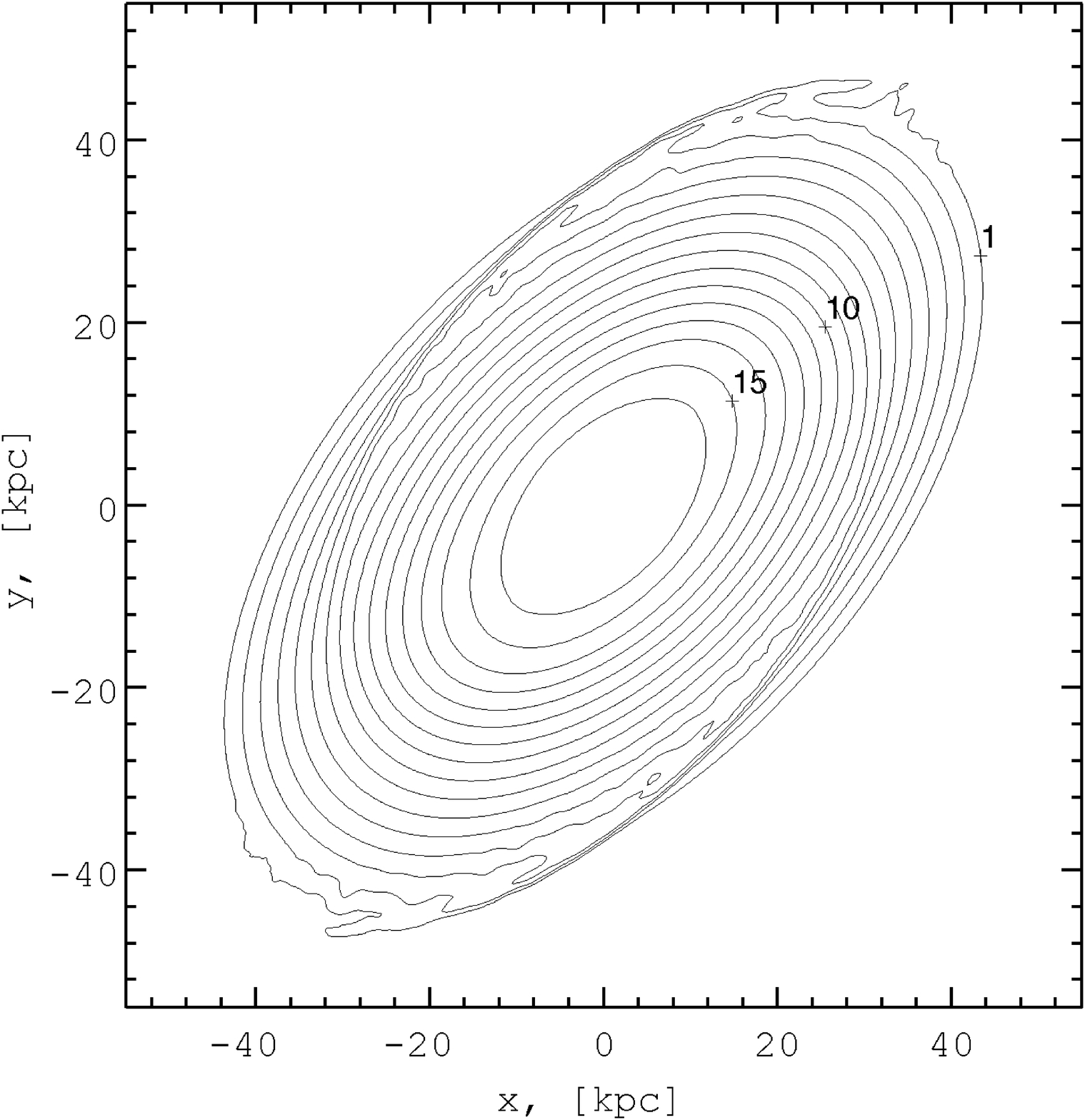}
\includegraphics[width=0.47\hsize]{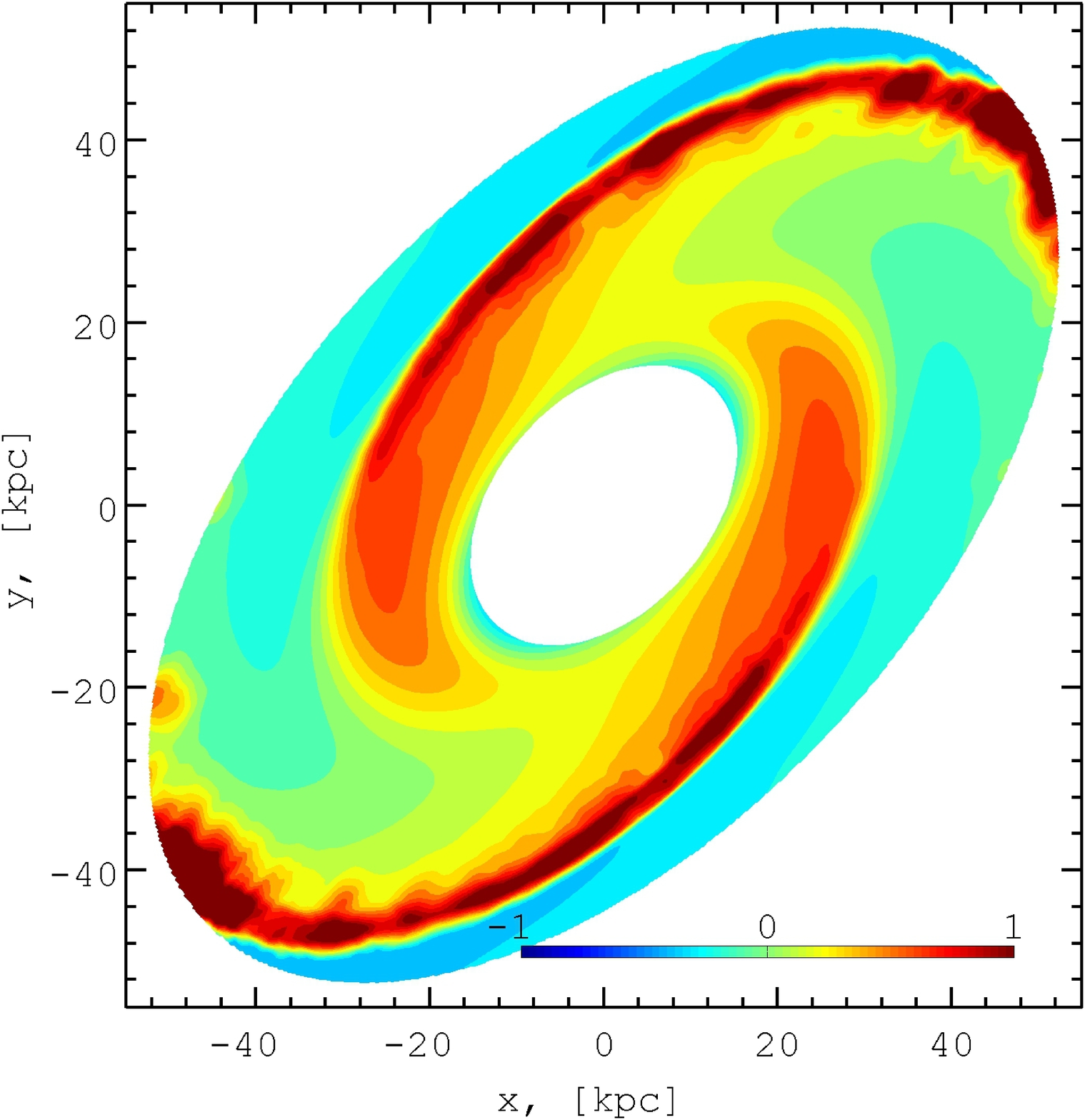} 
\includegraphics[width=0.47\hsize]{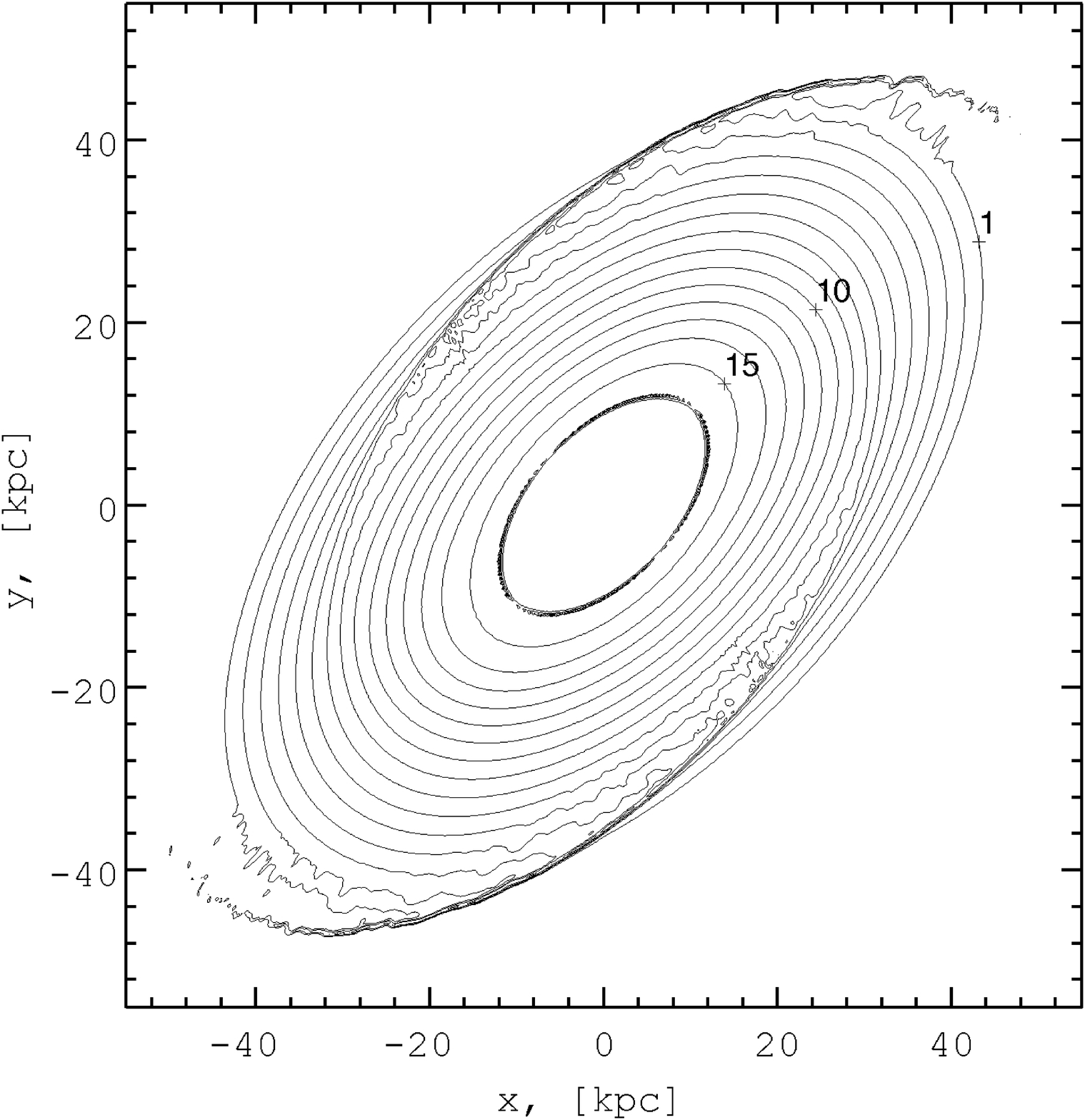}\hfill \includegraphics[width=0.47\hsize]{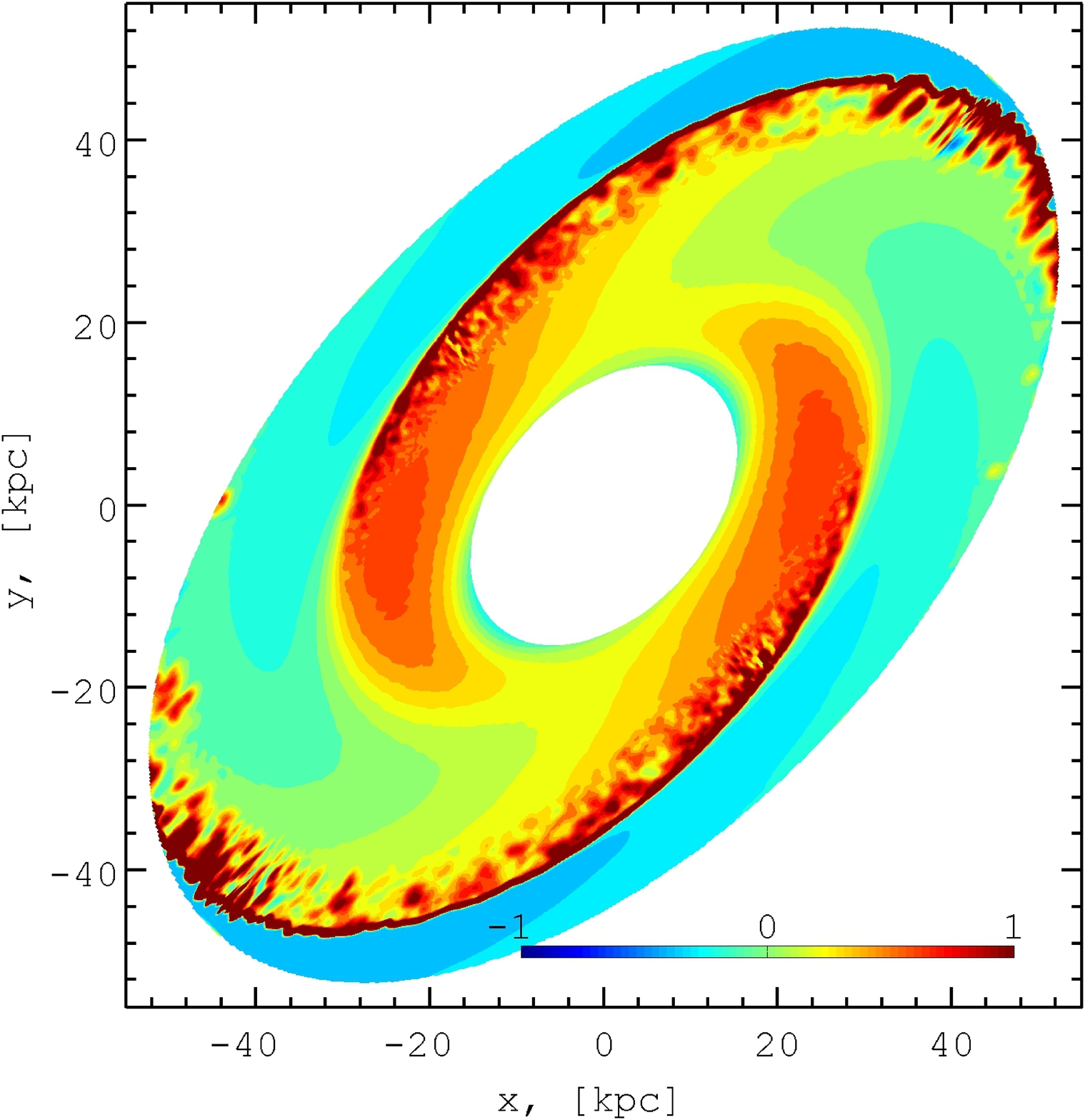}
\caption{Synthetic HI column density maps (left) for a simulated galaxy. Contours are in the range $(1 - 15) \times 10^{19}$~cm$^{-2}$.  The right frames show residual maps of the column density relative to the initial unperturbed state. The inner boundary disk~($r<6h = 18$~kpc) is masked by a white ellipse. Top row frames correspond to the  model J1 with the superposition of several modes  and small pitch angle; {\bf the middle and bottom} frames are for the model with a single $m=2$  more open spiral pattern  (model B7). Top and middle frames show the effect of a Gaussian filter adopted in the HI spectra calculation, bottom frames correspond to a calculation without beam smearing effects.}\label{fig::column_density}
\end{figure}

Channel maps of the simulated HI data cube are shown in Fig.~\ref{fig::hidatacub}. The HI data cube is smoothed down to a spatial
resolution\footnote{ For instance, in the paper by~\citet{2008A&A...490..555B}, the beam size for the velocity field is 22 arcsec (see Fig. 2 in that paper) where the conversion unit for 60 arcsec is 1.75 kpc (see Table 1 in the same paper). Thus we expect that for NGC~6946, 700 pc = 24 arcsec.} of 700~pc ($\approx0.2$~h) by using a Gaussian filter. The top panel illustrates the properties of the data cube for an unperturbed galactic disk (i.e., at the beginning of the simulation). The  bottom panel shows the synthetic data for the disk at $t=0.9~T_0$, that is, at a time when spiral structure is well established. Integration along the entire velocity range yields the observed gas density distribution~(see~Fig.~\ref{fig::column_density}).

For the unperturbed disk state the flux within each channel is characterized by regular structure~(see the top panel in Fig.~\ref{fig::hidatacub}). In contrast, the channel maps of the evolved disk show a complex structure, as determined by the presence of high-amplitude non-circular motions. In other words,  in the velocity channels the spectra show clear trace of the kinematic perturbations induced by the presence of prominent spiral structure. Features of this kind are often observed in grand-design spiral galaxies within the region occupied by the bright optical disk~\citep{1980A&A....88..149V,1981AJ.....86.1825B,2001A&A...370..765V}. Comparable kinematical studies of the outermost gaseous disk are quite rare. As an exception, we can refer to the paper by~\citet{2002AJ....123.3124F} where the properties of the extended, differentially rotating HI disk, spiral structure, and outer warp are described in great detail.  

 In the velocity dispersion the ``thermal contribution" $\sigma_{\rm th} = 3-5$~\kmps~(see Table~\ref{tab::tab1}) is comparable to or even higher than the contribution provided by the random motions $\sigma_{\rm random}$ in the inter-arm regions~(see Fig.~\ref{fig::line_profiles}). In contrast, in the vicinity of spiral arms the contribution of random motions is higher~(6-7~\kmps). We interpret  this feature as the result of shear flows and wiggle instability in the vicinity of spiral shocks, as noted at the end of Sect.~3.

\begin{figure}
\includegraphics[width=0.325\hsize]{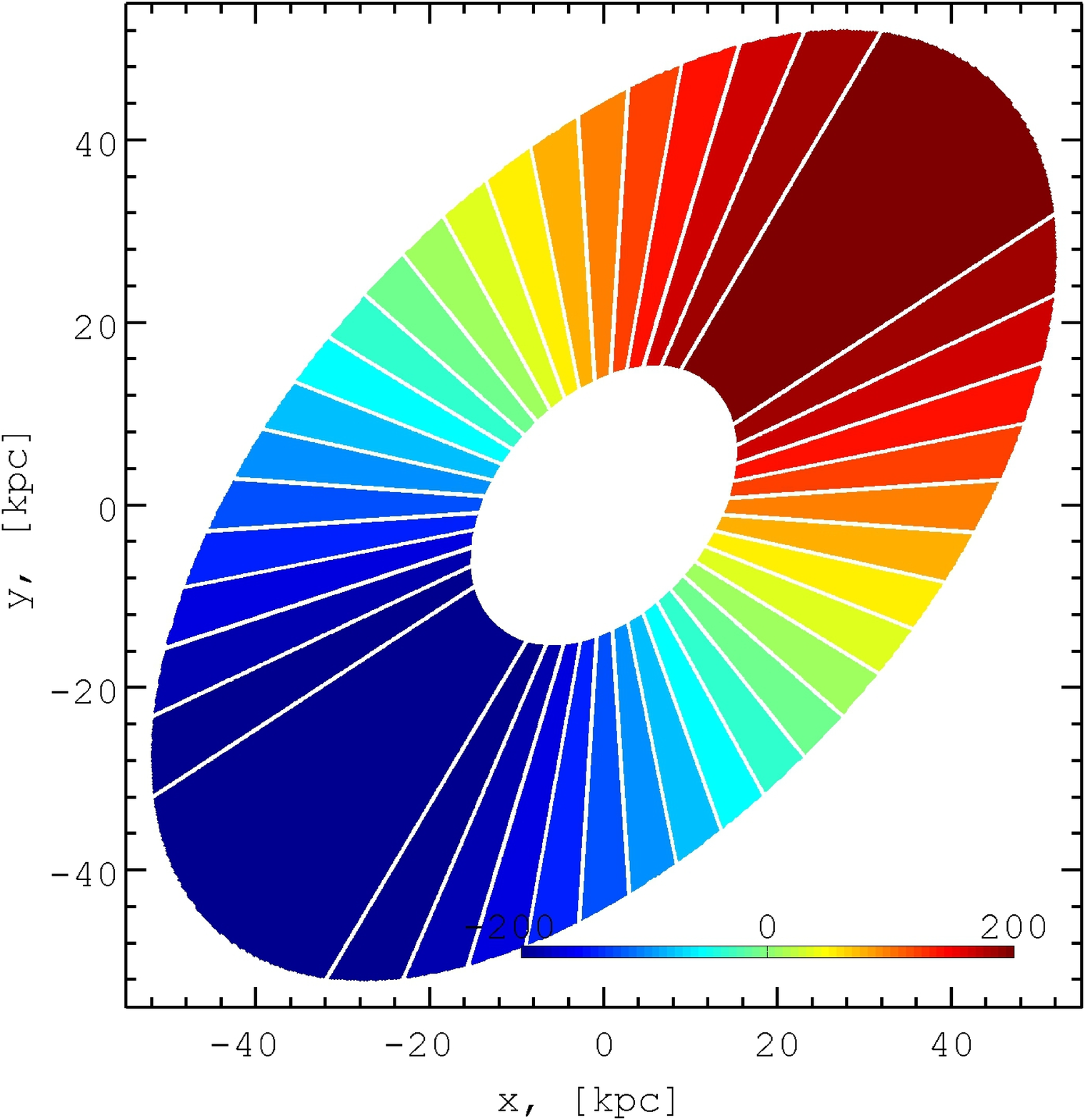}
\includegraphics[width=0.325\hsize]{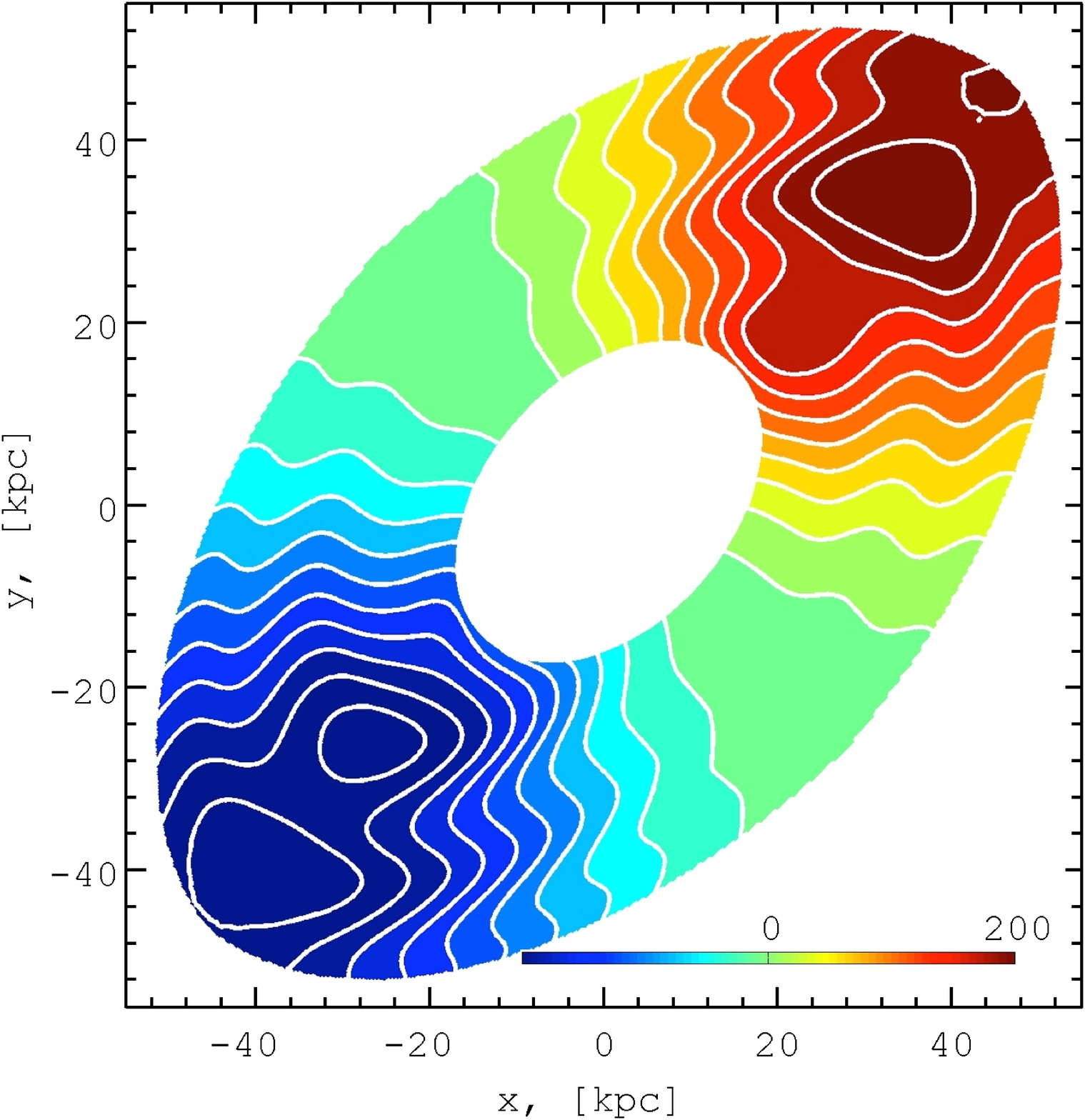}
\includegraphics[width=0.325\hsize]{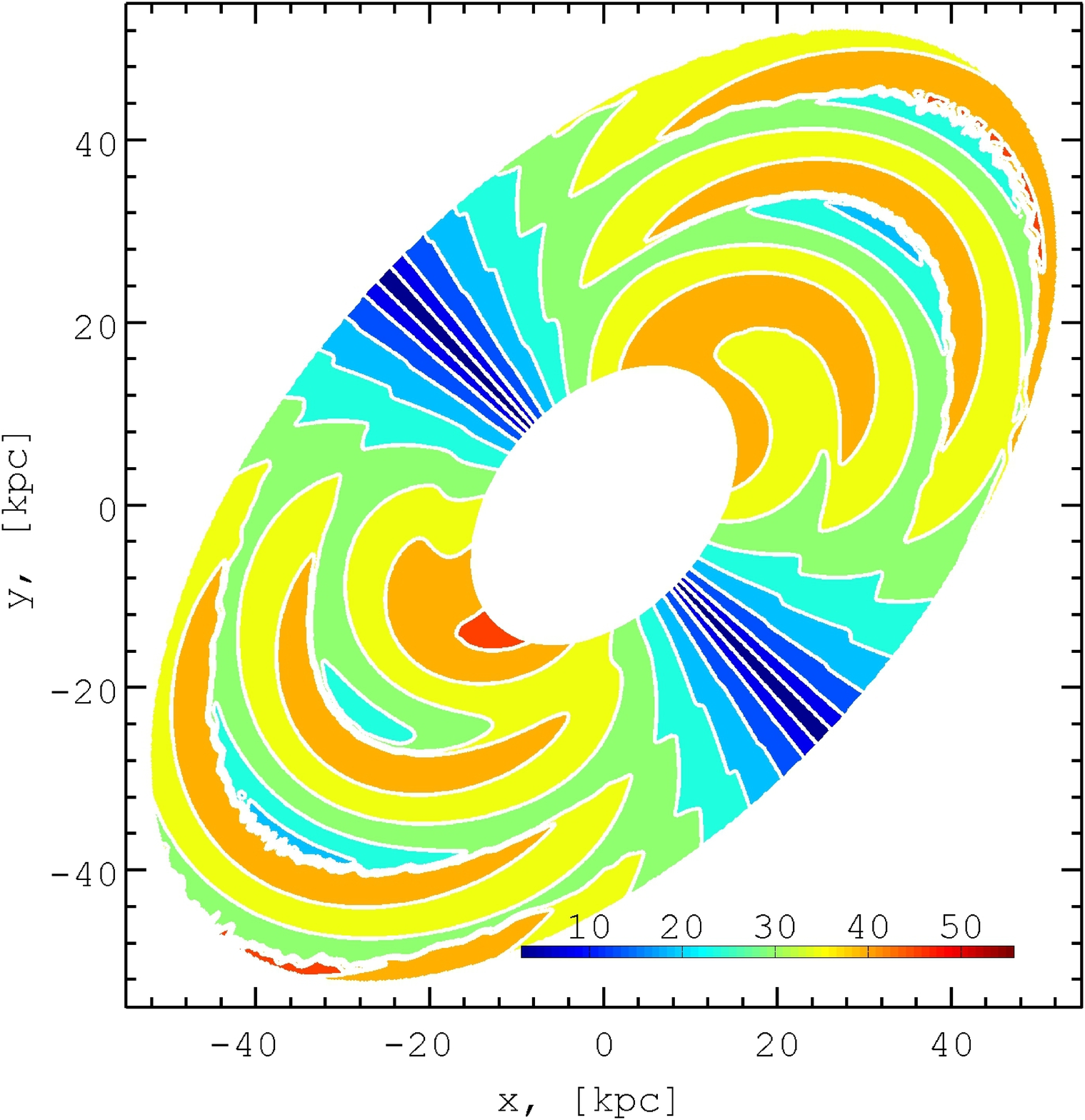} 
\includegraphics[width=0.325\hsize]{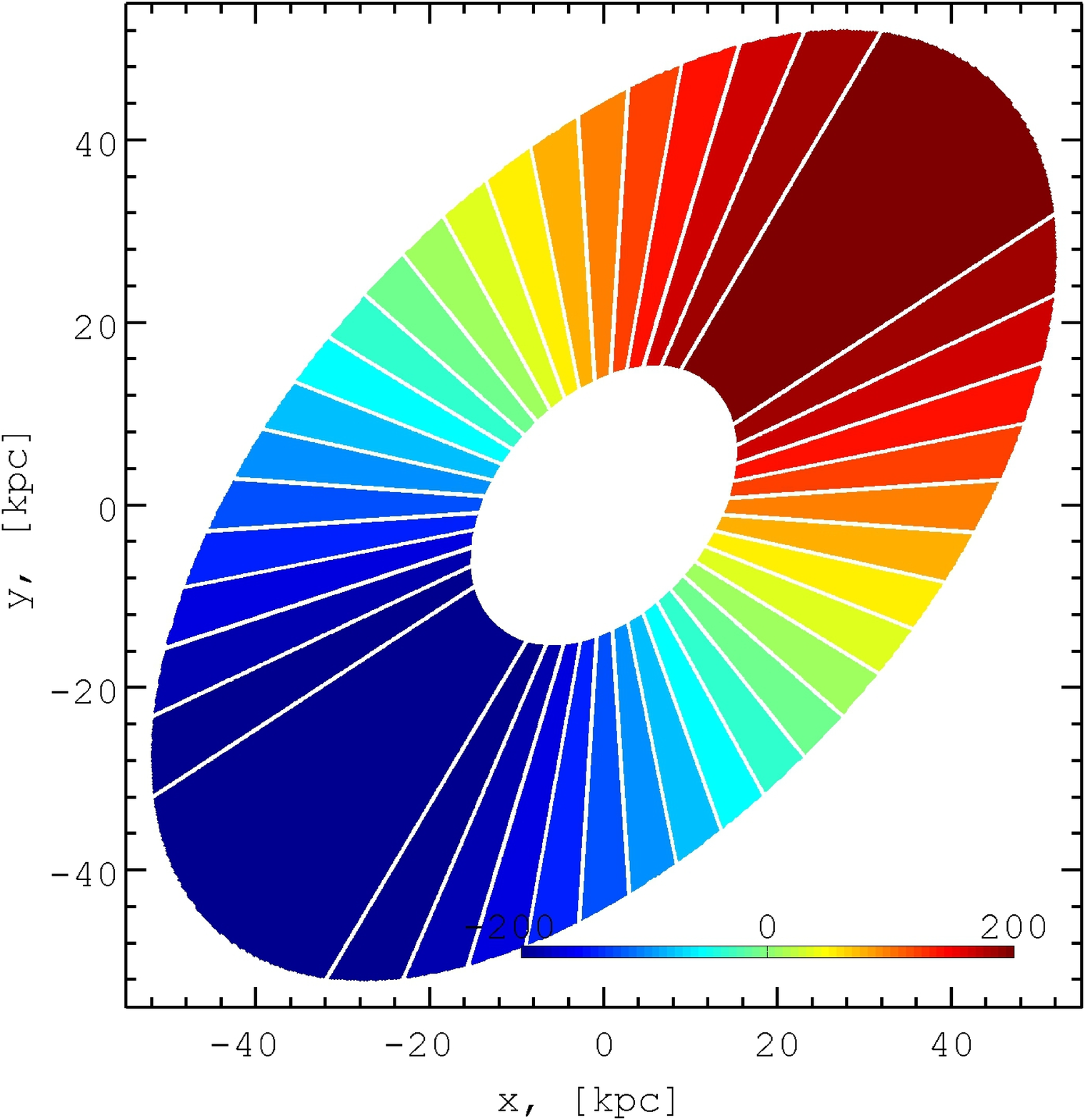}
\includegraphics[width=0.325\hsize]{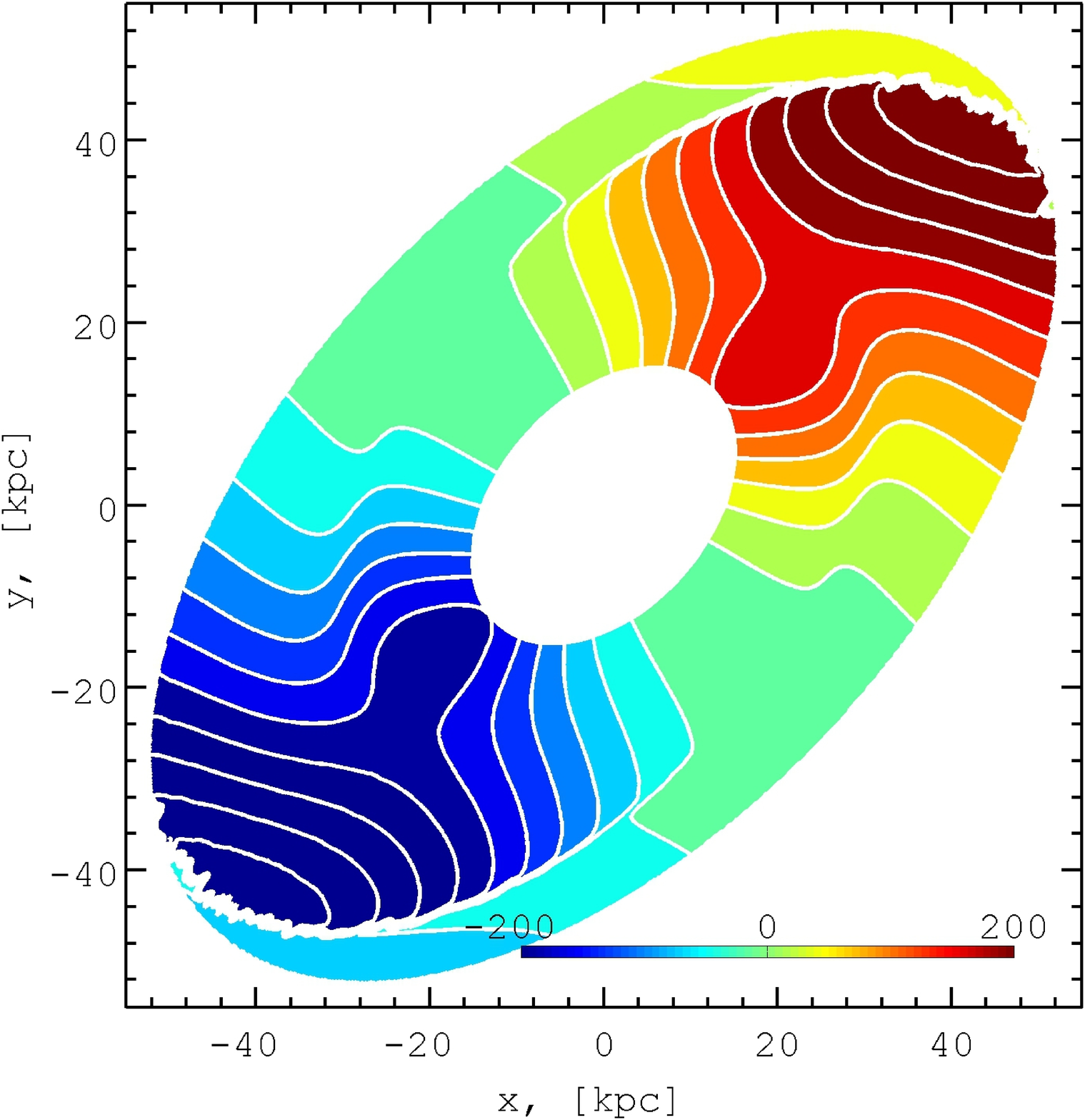}
\includegraphics[width=0.325\hsize]{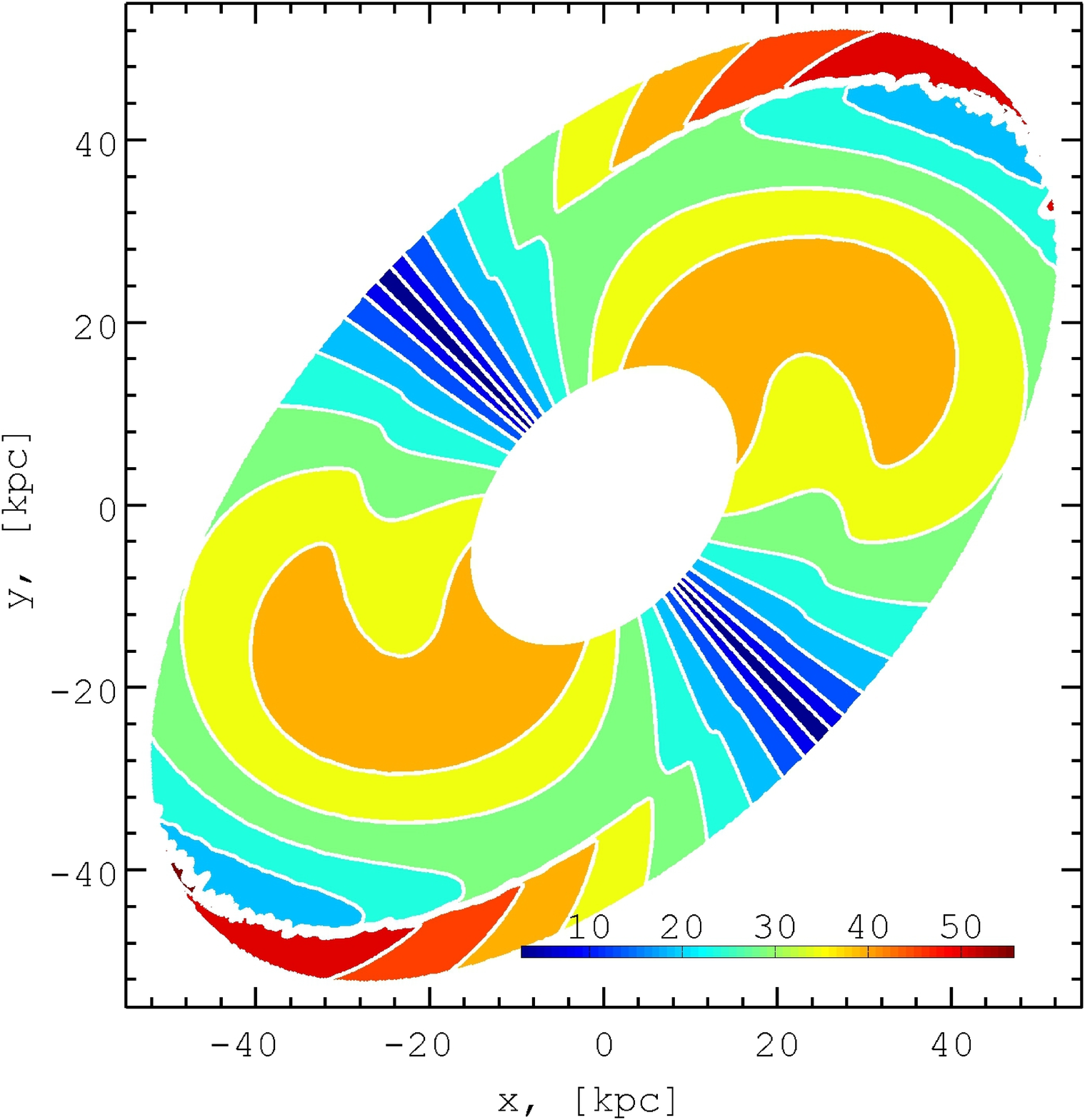} 
\caption{Synthetic LOS isovelocity contour maps for the initial state (left) and for the evolved state of the simulated gaseous disk~(middle) for two simulated galaxies. The right frames show residual maps of the LOS velocity distribution relative to the initial unperturbed state. Top frames correspond to the galaxy with the superposition of several modes and small pitch angle (model J1); the bottom frames are for the model with a single $m=2$  more open spiral pattern  (model B7). The inner boundary disk~($r<6h = 18$~kpc) is masked out.}\label{fig::2d_velocity}
\end{figure}

The resulting total HI column density map $N(x,y)$ for the evolved disk state is shown by means of black contours on the left frames of Fig.~\ref{fig::column_density}. Contours show the column density values in the range  $(0.1 - 20) \times 10^{19}$~cm$^{-2}$. Values of HI gas column density around $(0.05-0.1)\times 10^{19}$~cm$^{-2}$ can be detected by the Westerbork Synthesis Radio Telescope~\citep[e.g., see][for the outskirts of NGC 6946]{2008A&A...490..555B}, the Very Large Array~\citep[e.g., see][for the outskirts of NGC 2403]{2002AJ....123.3124F}, and the Giant Metrewave Radio Telescope~\citep[e.g., see][for NGC 3741, out to 38 optical scale lengths]{2005A&A...433L...1B}. We observe rather smooth spiral patterns, which cover the disk far outside the optical region. Residual maps of the column density distribution $(N(x,y) - N^0(x,y))/N(x,y)$ exhibit a clear spiral perturbation shape; here $N^0(x,y)$ is the column density distribution found from the synthetic spectral line observations of the initial unperturbed state.

At this stage, the rather sharp profiles of the line channel maps~(see Fig.~\ref{fig::hidatacub}) and 21-cm line profiles~(see Fig.~\ref{fig::line_profiles}) may appear to be not fully realistic. A more satisfactory comparison with the observations would be best performed on individual cases, for which the relevant instrumental parameters, telescope characteristics, observing mode, and specific detailed galaxy properties are all taken into proper account. Obviously, such thorough investigation would require a separate paper.

We also compare two images of the galactic disk seen in synthetic HI with and without beam smearing~(see Fig.~\ref{fig::column_density}). Clean images without smoothing show  very sharp density profiles across the spiral arms, which is obviously the effect of shock waves. Our simulations are sufficiently accurate to resolve shock fronts and the following formation of spurs. However, such features could be missed in radio observations because of low spatial resolution. We demonstrate this in  the middle and bottom frames of Fig.~\ref{fig::column_density}.

The kinematics of the gaseous disk is dominated by the differential rotation in the dark matter halo potential. A quasi-isothermal dark matter halo supports a flat rotation curve, which is clearly seen in the LOS velocity map in the left frames of Fig.~\ref{fig::2d_velocity}, where we show the diagram with the straight lines for the unperturbed iso-velocity contours. In the presence of spiral perturbations~(see middle frames of~Fig.~\ref{fig::2d_velocity}) the classic 'spider'-like diagram can still be recognized. A striking feature noted here is that of the quasi circular contours along the major axis. Such features are not expected in galaxies with flat rotation curves. Therefore, rather high velocity field perturbations can produce such features in the vicinity of the spiral arms. A quantitative analysis of this phenomenon can be done by studying the LOS velocity along the major axis~(see below). In particular, the residual velocity map\footnote{ We have checked that the morphologies of our residual velocity maps are not in conflict with the arguments presented by~\citet{1993ApJ...414..487C}.} reflects the structure of the density wave spiral perturbations~(see right frames of~Fig.~\ref{fig::2d_velocity}).

\begin{figure}
\includegraphics[width=0.495\hsize]{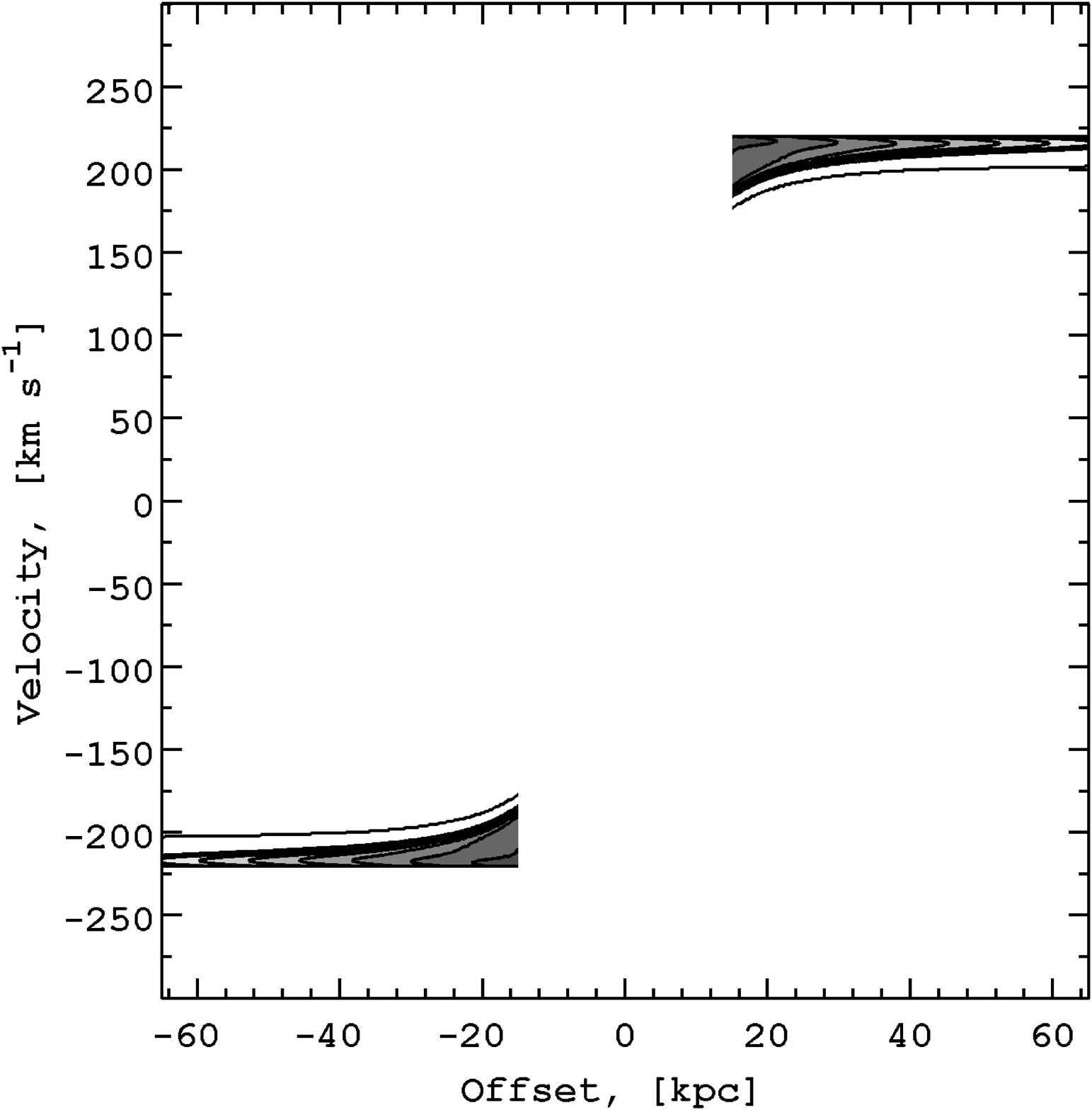}
\includegraphics[width=0.495\hsize]{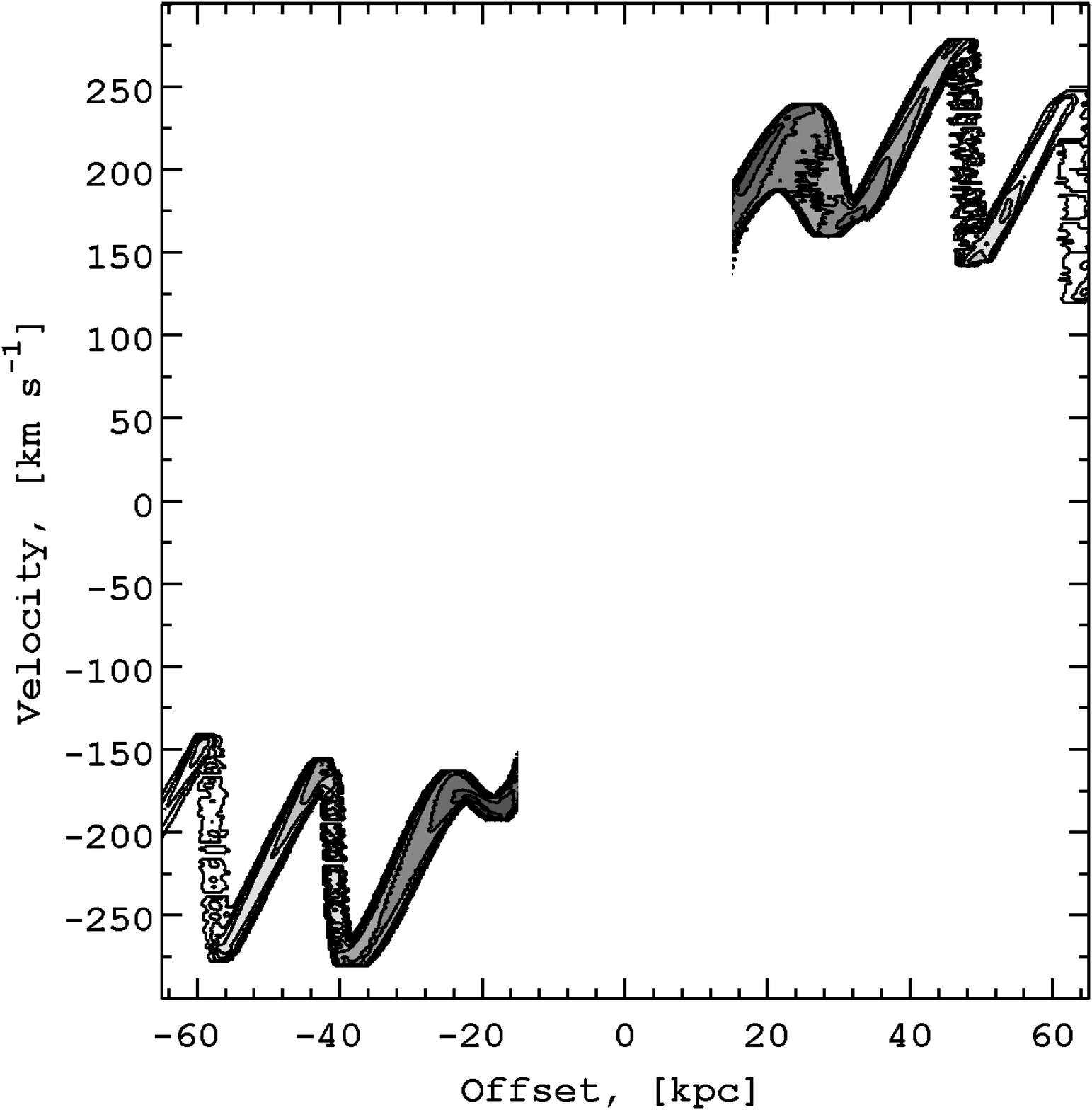}
\caption{Position - velocity diagram along the major axis~(corresponding  to the top frames of Fig.~\ref{fig::2d_velocity}): left --- for the initial unperturbed state, right --- for the evolved state of the simulated gaseous disk.  Contours are at 0.5, 1, 2, 3, 4, 5, 6, 7, 9, and 10 in units of $10^{19}$~cm$^{-2}$. The inner boundary disk~($r<6h = 18$~kpc) is masked out. }\label{fig::1d_velocity}
\end{figure}

The flat rotation curve for the basic~(initial) state is well established by the LOS velocity distribution along the major axis~(see Fig.~\ref{fig::1d_velocity}). There is also a significant scatter of the LOS velocity. From the dynamical point of view such scatter results from the effects of thickness and asymmetric drift present in the outer disk. For the perturbed~(evolved) disk state the flat rotation curve is not so clearly evident because non-linear spiral perturbations strongly affect the velocity field: rather regular variations of the LOS velocity, with amplitude $30-50$~\kmps, mark the location of the spiral arms. Such systematic variations, if found in 21~cm observations, would add strong support to the dynamical scenario we are proposing~\citep{2010A&A...512A..17B,2015MNRAS.451.2889K}. The position-velocity diagrams displayed in Fig.~\ref{fig::1d_velocity} exhibit features that are thinner than those observed in typical HI data: this suggests that, in the present model,  the effects leading to line broadening are not yet included in a satisfactory way, as would be desired for a fully realistic description.

\section{Conclusions}
In this paper we have presented results of 3D~hydrodynamical simulations of the establishment of prominent large-scale spiral structure in the outermost gaseous disk in galaxies. In the adopted general picture, the region associated with the bright optical disk acts as an ``engine" capable of producing global spiral structure without the need for external excitation; that is, the main disk is subject to few global spiral modes that are self-excited by the transfer of angular momentum to the outer regions by means of short trailing density waves. In the gaseous component, such outgoing density waves can penetrate through the outer Lindblad resonance and propagate outwards, as far as the gaseous disk extends. We have investigated the case in which a stationary disturbance at the inner boundary of the outermost gaseous disk mimics the situation addressed by the linear analysis of~\citet{2010A&A...512A..17B}. Here we have described the synthetic 21~cm dataset thus obtained for simulated galaxies. Our dynamical model points to the existence of several significant kinematical features related to the presence of spiral pattern perturbations that should be found in deep HI observations of the galactic outskirts. 

The results that we have obtained can be summarized as follows:

(i) With the inner boundary acting as a source of density waves, after a relatively rapid initial transient, a quasi-stationary spiral structure is established over the entire gaseous outer disk, well outside the bright optical disk. The amplitude of spiral structure is found to increase rapidly with radius. The properties of the spiral arms relatively close to the inner boundary follow the predictions of the linear theory~\citep{2010A&A...512A..17B}. Beyond $\approx 2$ optical radii, the supersonic motion of the perturbations relative to the basic state produces nonlinear shocks that become unstable with respect to shear-induced instabilities. 

(ii) Here we have focused on the predicted properties of the HI emission for selected velocity channels by considering simulated $21$-cm observations of a disk inclined at an angle of $30^\circ$. The synthetic spectra derived from the simulations presented in this paper suggest that the imprint of prominent non-axisymmetric structure of HI emission in the galactic outskirts could be identified and tested quantitatively in the relevant data cube of $21$-cm of radioastronomical observations.  In particular, the position-velocity diagram along the major axis suggests significant systematic variations of the LOS velocity~(up to 30-50~\kmps) at the galactic outskirts that should be detected in deep HI observations. Beam smearing can affect strongly the sharpness of the observed spiral waves in the HI gas; however, the large-scale morphology of the galactic outskirts should not depend significantly on smoothing effects.
\\
\\
We thank the referees for their detailed comments on the manuscript that helped to improve the quality of the paper. We wish to thank Alexei Moiseev and Evgenii Vasiliev for several helpful comments. SAK has been supported by a postdoctoral fellowship sponsored by the Italian MIUR. The numerical simulations have been performed at the Research Computing Center~(Moscow State University) under Russian Science Foundation grant~(14-22-00041) and Joint Supercomputer Center~(Russian Academy of Sciences). This work was supported by MIUR, by the President of the RF grant (MK-4536.2015.2) and RFBR grants~(13-02-00416, 14-22-03006, 15-02-06204, 15-32-21062). 

\bibliographystyle{jpp}

\bibliography{jpp-instructions}

\end{document}